\documentclass[a4paper,11pt]{article}
\usepackage{jheppub} 
\usepackage{lineno}
\usepackage{amsmath}
\usepackage[usenames,dvipsnames]{xcolor}
\usepackage[normalem]{ulem}
\usepackage{tikz}
\usepackage{comment}
\usepackage{enumerate}
\usepackage[utf8]{inputenc}
\usepackage{graphicx}
\usepackage{float}
\usepackage{todonotes}

\title{\boldmath Coarse Graining Holographic Black Holes in Higher Curvature Gravity}

\author{Qiongyu Qi}
\affiliation{Institute for Theoretical Physics, University of Amsterdam,\\
Science Park 904, 1098 XH, Amsterdam, The Netherlands}

\emailAdd{qiongyu.qi@student.uva.nl}

\abstract{We consider the holographic description of the dynamical black hole entropy in $f(R)$ higher curvature gravity which proposed by Hollands-Wald-Zhang. On the bulk side, we show that the coarse-grained entropy (outer entropy) of a generalized marginally trapped surface corresponds precisely to the Wald entropy associated with this surface. To get this result, we first formulate the AdS/CFT correspondence in the Einstein frame and derive the correspondence between the von Neumann entropy in the Einstein frame and the $f(R)$ frame. This facilitates the derivation of the correspondence between the two outer entropies in the two frames. Furthermore, we directly derive a focusing theorem associated with generalized expansion in $f(R)$ gravity. We then formulate how to construct the stationary null hypersurface for the generalized expansion and the junction condition to glue a hypersurface in $f(R)$ gravity. Combining these results, we directly derive the expression for the outer entropy in the $f(R)$ frame and identify its holographic dual.}

\begin{document}
\maketitle
\flushbottom

\section{Introduction}

In the non-dynamical regime, black hole entropy is conventionally governed by the area law (Bekenstein-Hawking entropy) \cite{Hawking:1975vcx}, 
\begin{equation}
    S_{\rm BH}=\frac{\text{Area}[\text{horizon}]}{4G}.
\end{equation}
This prescription was elegantly extended to higher-curvature theories by Iyer and Wald \cite{Wald:1993nt, Iyer:1994ys}, who identified the entropy as a Noether charge localized on the bifurcation surface $\mathcal{B}$. In particular, for $f(R)$ gravity
\begin{equation}
    S_{\rm Wald}=\frac{\int_{\mathcal{B}}f'(R)}{4G}.
\end{equation}
Following Iyer and Wald's attempt to generalize the Noether charge formalism to dynamical black holes, an entropy formula for higher curvature theories ($f(\rm Riemann)$ gravity) was derived by Wall \cite{Wall:2015raa}
\begin{equation}
    S_{\rm Wall}=-8\pi\int_{\mathcal{C}(v)}\Big(\frac{\partial L}{\partial R_{uvuv}}-4\frac{\partial^{2}L}{\partial R_{uiuj}\partial R_{vkvl}}K_{ij(u)}K_{kl(v)}\Big).
\end{equation}
Where $\mathcal{C}(v)$ is a cross section of the horizon, $K_{ij(a)}$ is the extrinsic curvature of the horizon in the $a$ direction. Remarkably, the Wall entropy matches to the holographic entanglement entropy computed by Dong \cite{Dong:2013qoa} for higher curvature gravity. Since the Dong and Wall entropies are derived in completely different contexts, it is not well understood whether their agreement is a coincidence or whether it holds more generally.

However, extending this framework to truly dynamical settings remains a non-trivial task. Away from the bifurcation surface, the evolution of the horizon introduces fundamental challenges in defining a unique, physical entropy. Two decades after the attempt of Iyer-Wald for generalizing the Noether charge formalism to dynamical black holes \cite{Wald:1993nt, Iyer:1994ys}, significant progress in approaching the dynamical black hole entropy was made in \cite{Hollands:2024vbe,Visser:2024pwz}. By applying the improved Noether charge method, a new notion of dynamical black hole entropy in general diffeomorphism-invariant theories of gravity under first order non-stationary perturbation, denoted by $S_{\text{dyn}}$, was introduced:
\begin{equation}
S_{\text{dyn}}=\frac{2\pi}{\kappa}\int_{\mathcal{C}}\tilde{\mathbf{Q}}_{\xi}=\frac{2\pi}{\kappa}\int_{\mathcal{C}}(\mathbf{Q}_{\xi}-\xi\cdot\mathbf{B}_{\mathcal{H}^{+}})\,,
\end{equation}
where ${\mathbf{Q}}_{\xi}$ is the Noether charge of the horizon Killing field $\xi^a$, and $\mathbf{B}_{\mathcal{H}^{+}}$ is a specific $(n-1)$-form constructed on the future horizon $\mathcal{H}^{+}$. In Einstein gravity, this entropy takes the form $S_{\text{dyn}} = (1 - v\partial_v)S_{\text{BH}}$\footnote{This formula  was also provided in \cite{Rignon-Bret:2023fjq} along with an alternative formula which vanishes on any cross section of a light cone in Minkowski spacetime.}, where $v$ is an affine parameter along the horizon. 

Significant progress about dynamical black hole entropy has also been made in various contexts, including black holes far from equilibrium \cite{Takeda:2024qbq, Ashtekar:2025qqa}, semi-classical interpretations of dynamical black holes \cite{ Bhattacharya:2025cld, Alonso-Serrano:2025lbo, Jia:2025tgf},  charged black holes \cite{Visser:2025jnf}, cosmology and dS spacetime \cite{Cai:2025ctb, Nojiri:2025gkq, Zhao:2025zny}, first order generalized expansions \footnote{We emphasize that the generalized here is the expansion for Wald entropy \eqref{definition of Generalized Expansion in section 1.2}, not the expansion for the generalized entropy \cite{Bekenstein:1973ur, Bousso:2015mna} used in quantum focusing theorem.} (or entropic expansions) and generalized focusing theorems \cite{Furugori:2025pmn, Kong:2024sqc, Yan:2024gbz}, and dynamical black
holes in the Einstein frame \cite{Kong:2024sqc}.

In this paper, we will propose a holographic interpretation of dynamical black holes entropy in $f(R)$ gravity. To do this, we will generalize the coarse grained black hole entropy proposed by Engelhardt and Wall \cite{Engelhardt:2017aux, Engelhardt:2018kcs} to $f(R)$ gravity. For these theories, we demonstrate that the dynamical entropy is $S_{\text{dyn}} = (1 - v\partial_v)S_{\text{Wald}}$. In particular, in $f(R)$ gravity, it has been shown in \cite{Kong:2024sqc, Jia:2025tgf, Furugori:2025pmn} that $S_{\text{dyn}}$ for $f(R)$ gravity equals $S_{\rm Wald}$ of the generalized apparent horizon. Motivated by these developments, we generalize the holographic coarse-grained black hole entropy to $f(R)$ gravity; we demonstrate that the outer entropy of the generalized minimar surface $\mu$ is equal to the Wald entropy of $\mu$
\begin{equation}
    S^{\text{(outer)}}_{f}[\mu]=\frac{\int_{\mu}f'(R)}{4G},
    \label{finally result in section 1}
\end{equation}
and find its boundary dual which is the simple entropy. This agrees with $S_{\rm dyn}$ in $f(R)$ gravity, and we will show that both  the outer entropy and the simple entropy  satisfy the second law. Therefore, it is natural to interpret the outer entropy and the simple entropy as the holographic interpretations of $S_{\rm dyn}$ in $f(R)$ gravity. This construction is non-perturbative in the classical gravity (valid, at minimum, to all orders in perturbation theory near equilibrium).

Since the matter fields will satisfy the Null Curvature Condition (NCC) in the Einstein frame if they satisfy the Null Energy Condition (NEC) in the $f(R)$ frame\footnote{We use the $f(R)$ frame to denote the frame of $f(R)$ gravity, see section \ref{Assumptions, Conventions, and Definitions}.}, the focusing theorem will be satisfied in the Einstein frame. Therefore, we will derive the outer entropy in the Einstein frame. It follows that we can evaluate the outer entropy in the $f(R)$ frame by the von Neumann entropy correspondence between the Einstein frame and the $f(R)$ frame.  We derive a nonlinear Raychaudhuri equation for the generalized expansion \eqref{definition of Generalized Expansion in section 1.2} which will guarantee the focusing theorem in $f(R)$ gravity, we will use this equation to derive the outer entropy directly in the $f(R)$ frame. This focusing theorem in $f(R)$ gravity will also help us define the boundary dual of the outer entropy.

This paper is organized as follows. In section \ref{section Einstein Frame}, we derive the correspondence between the Einstein frame and the $f(R)$ frame in asymptotic AdS spacetime and discuss the NCC condition and  junction conditions in the Einstein frame. In section \ref{section Coarse Gaining Entropy of Apparent Horizon in Einstein Frame}, we  derive the outer entropy in the Einstein frame. In section \ref{Relation Between the Outer Entropy in Einstein Frame and f(R) frame}, we will show the von Neumann entropy correspondence between the Einstein frame and the $f(R)$ frame, and using this correspondence show the result \eqref{finally result in section 1}. In section \ref{Deriving the Results in f(R)}, we  derive the focusing theorem in $f(R)$ gravity and directly show \eqref{finally result in section 1} in the $f(R)$ frame. Finally, in section \ref{The Boundary Dual of the Outer Entropy}, we  show that the simple entropy is the boundary dual of the outer entropy and discuss the second law.

\subsection{Assumptions, Conventions, and Definitions}
\label{Assumptions, Conventions, and Definitions}

This section establishes assumptions, conventions, and definitions that will be used throughout the paper.

\paragraph{$f(R)$ Frame.} In order to distinguish from the Einstein frame and for the convenience of subsequent descriptions, we use the $f(R)$ frame to denote the frame in which the gravitational Lagrangian is $f(R)$. We use the superscripts $E$ and $f$ to emphasize which frame we are working in.

\paragraph{Generalized Expansion.} Let $N_{k}$ be a null hypersurface with generating vector field $k^{a}$. For $f(R)$ gravity, we define the generalized expansion $\Theta_{k}$
\begin{equation}
\Theta_{k}=\theta_{k}+k^{a}\nabla_{a}\text{log}f'(R),
\label{definition of Generalized Expansion in section 1.2}
\end{equation}
Here $\theta_{k}$ is the ordinary expansion (details in section \ref{section 5.1}). Assume that $k^{a}=(\partial_{v})^{a}$ is affinely parametrized and $v$ is the affine parameter. We can rewrite it as
\begin{equation}
\Theta_{k}=\Theta_{v}=\partial_{v}\text{log}\big(\sqrt{\gamma}f'(R)\big).
\end{equation}

\paragraph{Generalized Extremal Surface.} A surface $X$ is generalized extremal if the generalized expansions of the two null orthogonal congruences fired from it both vanish:
\begin{equation}
\begin{split}
    \Theta_{l}&=0\\\Theta_{k}&=0.
    \end{split}
\end{equation}

\paragraph{Generalized Marginally Trapped Surface.} A surface is generalized marginal if the generalized expansion of the two null orthogonal congruences fired from $\mu$ satisfy:
\begin{align}
\begin{split}
    \Theta_{l}&<0\\\Theta_{k}&=0.
    \end{split}
\end{align}
\paragraph{Generalized Minimar Surface.} A generalized marginal surface $\mu$ will be called a generalized minimar surface if it additionally satisfies the following criteria:
\begin{enumerate}
    \item $\mu$ is homologous to $B$ (a entire connected component of the CFT at one time), and there exists a Cauchy slice $\Sigma_{min}[\mu]$ of the outer wedge $O_{W}^{f}[\mu]$ on which $\mu$ is a minimal Wald entropy surface homologous to $B$.
    \item There exists a choice of normalization for $l^{a}$ such that $\nabla_{k}\Theta_{l}=k^{a}\nabla_{a}\Theta_{l}\leq0$ on $\mu$,  with equality allowed only if $\Theta_{l}=0$ everywhere on $\mu$.
\end{enumerate}

\section{AdS/CFT in the Einstein Frame}
\label{section Einstein Frame}

In this section, we will assume the AdS/CFT correspondence in the original spacetime, and we work in the large-$N$, large-$\lambda$ limit, in which the bulk $\mathcal{M}$ is well-approximated by classical gravity \cite{Maldacena:1997re, Witten:1998qj}. We aim to demonstrate that if the spacetime is asymptotically AdS in the $f(R)$ frame, it remains asymptotically AdS in the Einstein frame. Furthermore, we show that matter fields satisfying the Null Energy Condition (NEC) in the $f(R)$ frame imply the satisfaction of the Null Curvature Condition (NCC) in the Einstein frame. Finally, we will give the junction condition for gluing spacetime regions across a codimension-2 surface.

\subsection{AdS/CFT Correspondence in the Einstein Frame}
Consider the action $I_{f}$ for $f(R)$ gravity in the $f(R)$ frame, given by
\begin{equation}
    I_{f}=\frac{1}{16\pi G}\int d^{D}x \sqrt{-g}\big( f(R)-2\Lambda\big)+I_{\text{matter}}.
\end{equation}
The modified Einstein equation in $f(R)$ gravity is 
\begin{equation}
    f'(R)R_{ab}-\frac{1}{2}g_{ab}f(R)+(g_{ab}\nabla_{c}\nabla^{c}-\nabla_{a}\nabla_{b})f'(R)+g_{ab}\Lambda=8\pi GT_{ab},
    \label{EOM of f(R)}
\end{equation}
here the Latin indices $abc$ are the abstract indices. Now we assume that
\begin{equation}
    f'(R)>0,\ f''(R)\neq0
\end{equation}
since $f'(R)>0$ ensures the positivity of the effective gravitational constant $G_{\rm eff} = G/f'(R)$ and $f''(R)\neq0$ will make our $f(R)$ gravity different from the Einstein gravity \cite{Sotiriou:2008rp, Nojiri:2017ncd, Kong:2024sqc}. 

We introduce an auxiliary field $\Psi$ and modify the action of $f(R)$ gravity as follows:
\begin{equation}
    I_{E}=\frac{1}{16\pi G}\int d^{D}x\sqrt{-g}\big(f'(\Psi)R+f(\Psi)-\Psi f'(\Psi)-2\Lambda\big)+I_{\text{matter}}.
\end{equation}
The on-shell condition for the auxiliary field $\Psi$ is
\begin{equation}
    \Psi=R,
\end{equation}
reducing the modified action $I_E$ to the original action $I_{f}$.

We now proceed to the Einstein frame by introducing the Weyl transformation
\begin{equation}
    g_{ab}^{E}=\big(f'(\Psi)\big)^{\frac{2}{D-2}}g_{ab}.
    \label{Weyl transformation}
\end{equation}
One can easily verify that $\Omega^{2}=(f'(\Psi))^{\frac{2}{D-2}}>0$ (we assume $\Omega>0$). Furthermore, the Weyl transformation preserves the causal structure of the spacetime. We redefine the auxiliary field $\Psi$ by
\begin{equation}
    \psi=\frac{1}{\sqrt{16\pi G}}\sqrt{\frac{2(D-1)}{D-2}}\ \text{log}f'(\Psi).
    \label{definition of psi}
\end{equation}
We use the metric $g^{E}_{ab}$ and $\psi$ rewrite the gravitational action \cite{Whitt:1984pd, Barrow:1988xh, Kong:2024sqc}
\begin{equation}
    S_{f}=\int d^{D}x\sqrt{-g^{E}}\Big(\frac{R^{E}}{16\pi G}-\frac{1}{2}g^{E}_{ab}\partial^{a}\psi\partial^{b}\psi-V_{0}(\psi)-\frac{2}{16\pi G}\big(f'(\Psi)\big)^{-\frac{D}{D-2}}\Lambda\Big)+S_{\text{matter}},
\end{equation}
where
\begin{equation}
    V_{0}(\psi)=\frac{\big(f'(\Psi)\big)^{-\frac{D}{D-2}}}{16\pi G}\Big(\Psi f'(\Psi)-f(\Psi)\Big).
\end{equation}

Suppose in the $f(R)$ frame, the on-shell asymptotic AdS curvature is $R_{0}$. From \eqref{EOM of f(R)}, we can determine the vacuum solution
\begin{equation}
    f'(R_{0})r_{0}-\frac{D}{2}f(R_{0})+D\Lambda=0.
\end{equation}
One can show that, for the on-shell action, the potential and cosmological terms in the Einstein frame combine to form a new cosmological constant $\Lambda'$:
\begin{equation}
    V_{0}(R_{0})+\frac{2}{16\pi G}\big(f'(R_{0})\big)^{-\frac{D}{D-2}}\Lambda=\frac{2}{16\pi G}\Big(\frac{D-2}{2D}f'(R_{0})^{-\frac{2}{D-2}}R_{0}\Big)=\frac{2}{16\pi G}\Lambda'.
    \label{Lambda'}
\end{equation}
To obtain the standard form of the action in the Einstein frame, we define a new potential $V(\psi)$
\begin{equation}
    V(\psi)=V_{0}(\psi)+\frac{2}{16\pi G}\big(f'(\Psi)\big)^{-\frac{D}{D-2}}\Lambda-\frac{2\Lambda'}{16\pi G}.
\end{equation}
The action $I_E$ can then be rewritten in the Einstein frame as
\begin{equation}
    I_{E}=\int d^{D}x\sqrt{-g^{E}}\Big(\frac{R_{E}-2\Lambda'}{16\pi G}-\frac{1}{2}g^{E}_{ab}\partial^{a}\psi\partial^{b}\psi-V(\psi)\Big)+I_{\text{matter}}.
\end{equation}
This ensures that the new potential $V(\psi)$ vanishes asymptotically near the AdS boundary. Finally, to ensure the field satisfies a well-defined boundary condition, we define a shifted field $\phi$:
\begin{equation}
    \phi=\psi-\frac{1}{\sqrt{16\pi G}}\sqrt{\frac{2(D-1)}{D-2}}\ \text{log}f'(R_{0}).
\end{equation}
This ensures that the field $\phi$ vanishes asymptotically near the AdS boundary. Finally, the action in the Einstein frame should be written as 
\begin{equation}
     I_{E}=\int d^{D}x\sqrt{-g^{E}}\Big(\frac{R_{E}-2\Lambda'}{16\pi G}-\frac{1}{2}g^{E}_{ab}\partial^{a}\phi\partial^{b}\phi-V(\phi)\Big)+I_{\text{matter}}.
     \label{final action}
\end{equation}
Asymptotically, the field $\phi$ and the potential $V(\phi)$ satisfy the following boundary conditions:
\begin{equation}
    \lim\phi\to0,\ \lim V(\phi)\to 0.
    \label{boundary condition of phi}
\end{equation}

Now, from the above constructions, we can derive a correspondence between on-shell asymptotic AdS solutions in the $f(R)$ frame and the Einstein frame. On the one hand, every on-shell asymptotic AdS solution in the $f(R)$ frame can be translated into an asymptotic AdS solution in the Einstein frame, but the spacetime satisfies the Einstein equation \eqref{final action} with an additional field $\phi$ satisfying the boundary condition \eqref{boundary condition of phi}. On the other hand, every on-shell solution of the action \eqref{final action} with the same cosmological constant $\Lambda'$ under the boundary condition \eqref{boundary condition of phi} corresponds to an on-shell solution in the $f(R)$ frame via the following steps:
\begin{enumerate}
    \item Reconstructing $\psi$ by
    \begin{equation}
        \psi=\phi+\frac{1}{\sqrt{16\pi G}}\sqrt{\frac{2(D-1)}{D-2}}\ \text{log}f'(R_{0}),
        \label{inverse transformation 1}
    \end{equation}
    here $R_{0}$ can be obtained by \eqref{Lambda'}.
    \item Using the definition of $\psi$ \eqref{definition of psi}, we can get the conformal factor
    \begin{equation}
        f'(R)=\exp\Big(\sqrt{16\pi G\frac{D-2}{2(D-1)}}\psi\Big),
        \label{inverse transformation 2}
    \end{equation}
    here we use the on-shell condition $\Psi=R$.
    \item Finally, we use the conformal factor to get the geometry in the $f(R)$ frame
    \begin{equation}
        g_{ab}=\big(f'(R)\big)^{-\frac{2}{D-2}}g^{E}_{ab}.
        \label{inverse transformation 3}
    \end{equation}
\end{enumerate}
Thus, we obtain a one-to-one correspondence between on-shell asymptotic AdS solutions in the $f(R)$ frame and the Einstein frame.

As a consistency check, we consider the Weyl transformation of the Ricci scalar \cite{Wald:1984rg}
\begin{equation}
    R^{E}(g)=\Omega^{-2}\big(R-2(D-1)\nabla^{2}\text{log}\Omega-(D-1)(D-2)(\nabla\text{log}\Omega)^{2}\big).
\end{equation}
Consider the on-shell solution $\Psi=R$, when we approach the asymptotic boundary, $\Omega$ approaches a constant, while $\nabla^2\Omega$ and $\nabla\Omega$ vanish asymptotically. Therefore, if the original spacetime is asymptotic AdS with curvature $R_{0}$, the spacetime in the Einstein frame remains asymptotic AdS, and the asymptotic scalar curvature in the Einstein frame
\begin{equation}
    R^{E}_{0}=\big(f'(R_{0})\big)^{-\frac{2}{D-2}}R_{0}.
\end{equation}
This agrees with our previous result \eqref{final action}, since in the Einstein frame our theory is Einstein gravity, the asymptotic curvature in the Einstein frame
\begin{equation}
    R^{E}_{0}=\frac{2D}{D-2}\Lambda'=f'(R_{0})^{-\frac{2}{D-2}}R_{0}<0.
\end{equation}
Therefore, assuming the validity of the AdS/CFT correspondence in the Einstein frame is justified.

\subsection{NCC and Junction Conditions in the Einstein Frame}

In this subsection, we derive the Null Curvature Condition (NCC) in the Einstein frame
\begin{equation}
    R^{E}_{ab}k^{a}k^{b}\geq0,
\end{equation}
assuming the matter satisfies the Null Energy Condition (NEC) in the $f(R)$ frame. This is necessary because the derivation of the outer entropy requires the maximin construction \cite{Wall:2012uf} of HRT surface and the focusing theorem in the Einstein frame. Furthermore, junction conditions in the Einstein frame are required to glue spacetime regions when we derive the outer entropy in the Einstein frame.

\subsubsection{NCC Condition in the Einstein Frame}

Consider the NCC in the Einstein frame. The action is given by
\begin{equation}
     I_{E}=\int d^{D}x\sqrt{-g^{E}}\Big(\frac{R_{E}-2\Lambda'}{16\pi G}-\frac{1}{2}g^{E}_{ab}\partial^{a}\phi\partial^{b}\phi-V(\phi)\Big)+I_{\text{matter}}.
\end{equation}
Defining the stress tensor in the Einstein frame as $T_{ab}^E$, one finds that
\begin{equation}
T^{E}_{ab}=T^{\text{m}}_{ab}+\partial_{a}\phi\partial_{b}\phi-g^{E}_{ab}\Big(\frac{1}{2}g_{E}^{cd}\partial_{c}\phi\partial_{d}\phi+V(\phi)\Big),
\label{stress tensor in the Einstein frame}
\end{equation}
where 
\begin{equation}
T^{\text{m}}_{ab}=-\frac{2}{\sqrt{-g^{E}}}\frac{\delta I_{\text{matter}}}{\delta g_{E}^{ab}}=\frac{T_{ab}}{f'(R)}.
\label{relation between Tm and T}
\end{equation}
Consider a null vector $k_E^a$, we aim to show that the stress tensor in the Einstein frame satisfies the null energy condition
\begin{equation}
    T_{ab}^{E}k_{E}^{a}k_{E}^{b}\geq 0.
\end{equation}
Since the Weyl transformation preserves the causal structure, the null vector $k_{E}^{a}$ remains null in the $f(R)$ frame. Therefore, if the $f(R)$ frame satisfies the null energy condition, then
\begin{equation}
    T^{\text{m}}_{ab}k_{E}^{a}k_{E}^{b}=\frac{T_{ab}k_{E}^{a}k_{E}^{b}}{f'(R)}\geq 0,
\end{equation}
provided that $f'(R)>0$.

It remains to verify that the residual part of the stress tensor satisfies the null energy condition. Since $k_{E}^{a}$ is a null vector
\begin{equation}
    g^{E}_{ab}k^{a}_{E}k_{E}^{b}=0,
\end{equation}
we only need to ensure the kinetic term satisfies the NEC. It is easy to show that
\begin{equation}
k_{E}^{a}k_{E}^{b}\partial_{a}\phi\partial_{b}\phi=(k\cdot\partial\phi)^{2}\geq 0.
\end{equation}
Therefore, we find that the stress tensor in the Einstein frame satisfies the NEC
\begin{equation}
    T_{ab}^{E}k_{E}^{a}k_{E}^{b}\geq 0.
\end{equation}
Consider the equations of motion for gravity in the Einstein frame
\begin{equation}
    R^{E}_{ab}-\frac{1}{2}g^{E}_{ab}R^{E}+\Lambda'g^{E}_{ab}=8\pi GT^{E}_{ab}.
\end{equation}
It follows that
\begin{equation}
    R^{E}_{ab}k_{E}^{a}k_{E}^{b}=8\pi GT^{E}_{ab}k_{E}^{a}k_{E}^{b}\geq 0
\end{equation}
Thus, we have demonstrated that if the NEC holds in the $f(R)$ frame, the NCC is satisfied in the Einstein frame.

\subsubsection{Junction Conditions in the Einstein Frame}

Deriving the outer entropy in the Einstein frame requires gluing two spacetime patches through a codimension-2 surface $\sigma$, ensuring the resulting spacetime satisfies the Einstein equation and remains free of singularities at the gluing surface $\sigma$. Consequently, establishing the proper junction conditions is essential for the gluing procedure in the Einstein frame.

We define the discontinuity of a quantity $F$ across $\sigma$, traversing from $V_{\rm out}$ to $V_{\rm in}$, as:
\begin{equation}
    [F]=F|_{\sigma_{\text{out}}}-F|_{\sigma_{\text{in}}}.
\end{equation}

\paragraph{Codimension-2 junction conditions.} Define $V_{\text{out}}$ and $V_{\text{in}}$ as
\begin{equation}
    \begin{split}
V_{\text{out}}&=D[{\Sigma_{\text{out}}}[\sigma_{\text{out}}]]\\V_{\text{in}}&=D[\Sigma_{\text{in}}[\sigma_{\text{in}}]].
    \end{split}
\end{equation}
Here, $\sigma$ is a codimension-2 surface that partitions the Cauchy slice $\Sigma$ into $\Sigma_{\text{out}}$ and $\Sigma_{\text{in}}$, $\sigma_{\rm out}$ and $\sigma_{\rm in}$ denote the surface $\sigma$ embedded in $V_{\rm out}$ and $V_{\rm in}$, respectively. The regions $V_{\rm out}$ and $V_{\rm in}$ can be glued to one another with a finite stress-energy tensor under the following conditions \cite{Engelhardt:2018kcs}:

\begin{enumerate}
    \item The surfaces $\sigma_{\text{out}}$ and $\sigma_{\text{in}}$ are isometric and can thus be identified as a single surface $(\sigma,h)$, where $h$ is the induced metric of $\sigma$.
    \item There exists a choice of $k^{a}_{E}$ and $l^{a}_{E}$ defined on both side of $\sigma$ such that the following conditions are satisfied:
    \begin{equation}
        \begin{split}
            [\theta^{E}_{k_{E}}]&=0\\ [\theta^{E}_{l_{E}}]&=0\\ [\chi^{E}_{k_{E}}]&=-[\chi^{E}_{l_{E}}]=0.
        \end{split}
    \end{equation}
    \label{junction conditions in the Einstein frame}
\end{enumerate}
Here $\chi^{E}$ is the extrinsic twist potential. Then the null-null components of the stress tensor are finite, and the Einstein equation is distributionally well-defined.

\section{Coarse-Grained Entropy of Marginally Trapped Surfaces in the Einstein Frame}
\label{section Coarse Gaining Entropy of Apparent Horizon in Einstein Frame}

In the Einstein frame, consider a $(D-2)$-dimensional surface $\sigma$. The outer entropy associated with $\sigma$ (homologous to $B$, an entire connected component of the CFT) is defined by maximizing the von Neumann entropy over all possible inner wedge data $\{\alpha\}$, while keeping the outer wedge $O_W^E[\sigma]$ fixed \cite{Engelhardt:2018kcs}:
\begin{equation}
    S_{E}^{\text{(outer)}}[\sigma]=\max_{\{\alpha\}}[S_{\text{vN}}(\rho^{E\{\alpha\}}_{B})].
\end{equation}
According to the AdS/CFT correspondence in the Einstein frame, each allowed spacetime constructed by attaching an inner wedge $I_W^E[\sigma]$ corresponds to a boundary state $\rho_B^{E\{\alpha\}}$, with von Neumann entropy given by:
\begin{equation}
    S_{\text{vN}}(\rho_{B}^{E\{\alpha\}})=-\text{tr}(\rho_{B}^{E\{\alpha\}}\text{log}\rho_{B}^{E\{\alpha\}})=\frac{\text{Area}_{E}[X^{\{\alpha\}}]}{4G},
\end{equation}
where $X^{\{\alpha\}}$ is the HRT surface homologous to the boundary component $B$ (we will always take $B$ as one of the conformal boundary of the AdS black hole). In the Einstein frame, the stress tensor $T_{ab}^E$ satisfies the NEC, which implies the NCC (as the equations of motion are the Einstein equations). Together with some global assumptions, we can use the maximin construction to define the HRT surface \cite{Wall:2012uf}: we first find the surface $\min(B,\Sigma)$ homologous to $B$ that minimizes the area on a given Cauchy slice $\Sigma$. We then choose $\Sigma$ to maximize the area of this minimal surface over all possible Cauchy slices.

We now demonstrate that the outer entropy of a marginally trapped surface $\mu$ (satisfying the minimar condition in Einstein frame\footnote{Minimar condition: 1. $\mu$ is homologous to a connected component $B$ of CFT at one time. And there exists a Cauchy slice $\Sigma$ in $O^{E}_{W}[\mu]$ such that $\mu$ has minimal area on $\Sigma$. 2. There exists a choice of normalization for $l^{a}_{E}$ such that $\nabla^{E}_{k_{E}}\theta^{E}_{l_{E}}\leq 0$ on $\mu$, with equality allowed only if $\theta^{E}_{l_{E}}=0$ everywhere on $\mu$.}) is proportional to its area in the Einstein frame. First, we show that the outer entropy of a marginally trapped surface $\mu$ is bounded from above by its area:
\begin{equation}
    S_{E}^{\text{(outer)}}[\mu]=\max \Big[S_{\text{vN}}(\rho_{B}^{E\{\alpha\}})=\frac{\text{Area}_{E}[X^{\{\alpha\}}]}{4G}\Big]\leq\frac{\text{Area}_{E}[\mu]}{4G}.
\end{equation}
Next, we demonstrate the existence of a configuration $\{\alpha\}$ and a spacetime $(\mathcal{M}^{\{\alpha\}}, g^E)$ for which the von Neumann entropy saturates this bound \cite{Engelhardt:2018kcs, Engelhardt:2017aux}; we will provide a simplified proof below.

\subsection{The Upper Bound of the Outer Entropy}
\label{The Upper Bound of the Outer Entropy}

We now show that $S_{E}^{\text{(outer)}}[\mu]$ is bounded by $\frac{\text{Area}[\mu]}{4G}$ in the Einstein frame. Assume the surface $\mu$ is marginally trapped. Let $N_{\pm k_{E}}(\mu)$ denote the null congruences in the Einstein frame generated from $\mu$ by the null geodesics $k_E^a$ and $-k_E^a$, where $k_E^a$ is future-directed. The representative of $\mu$ on Cauchy slice $\Sigma$ \cite{Wall:2012uf, Engelhardt:2018kcs, Engelhardt:2017aux}
\begin{equation}
    \overline{\mu}=N_{\pm k_{E}}\cap\Sigma,
\end{equation}
since $\overline{\mu}$ is homologous to $\mu$ and $B$, it is also homologous to the HRT surface $X^{\{\alpha\}}$. It follows from the NCC in the Einstein frame that the area of $\overline{\mu}$ is bounded by the area of $\mu$ \cite{Wall:2012uf}
\begin{equation}
    \text{Area}_{E}[\overline{\mu}]\leq\text{Area}_{E}[\mu].
\end{equation}
Under the NCC, the area of $\mu$ is maximal on $N_{\pm k_{E}}$.

Consider a spacetime $(\mathcal{M}^{\{\alpha\}}, g^E)$ with a fixed outer wedge $O_W^E[\mu]$. By the maximin construction \cite{Wall:2012uf}, there exists a Cauchy slice $\Sigma$ such that $X^{\{\alpha\}}$ is the minimal surface on $\Sigma$, therefore
\begin{equation}
    S_{\text{vN}}(\rho_{B}^{E\{\alpha\}})=\frac{\text{Area}_{E}[X^{\{\alpha\}}]}{4G}\leq\frac{\text{Area}_{E}[\overline{\mu}]}{4G}\leq\frac{\text{Area}_{E}[\mu]}{4G}.
\end{equation}
Since the outer entropy $S^{\text{(outer)}}[\mu]$ is the maximal von Neumann entropy, this implies
\begin{equation}
    S_{E}^{\text{(outer)}}[\mu]\leq\frac{\text{Area}_{E}[\mu]}{4G}.
\end{equation}

\subsection{Construct the Geometry that Saturates the Upper Bound}
\label{Construct the Geometry that Saturates the Upper Bound}

We now construct a spacetime, or equivalently an inner wedge $I^{E}_{W}[\mu]$ for which the von Neumann entropy $S_{\text{vN}}(\rho^{E}_{B})$ saturates the upper bound $\text{Area}[\mu]/4G$ in the Einstein frame. In this subsection, we will construct an extremal surface $X$ homologous to $B$ with the same area as the marginally trapped surface $\mu$ in the Einstein frame by gluing stationary null hypersurface $N_{-k}$. We then demonstrate the existence of an extremal surface $X$ on $N_{-k}$ and use CPT reflection across $X$ to construct the full spacetime\footnote{The geometry surfaces we construct form a null Cauchy slice $\Sigma=N'_{-l}\cup N_{-k}\cup N_{-l}$. The null Cauchy slice satisfies all the constraint equations and the junction condition. The whole spacetime $(\mathcal{M},g^{E})$ is defined by the Cauchy evolution of this slice (the characteristic initial data problem) \cite{Rendall1990ReductionOT, SAHayward_1993, Brady:1995na, Luk:2011vf}.} in the Einstein frame. In this spacetime, $X$ is the HRT surface, and this implies that the outer entropy of the marginally trapped surface in the Einstein frame is the Bekenstein-Hawking entropy of the marginally trapped surface \cite{Engelhardt:2018kcs, Engelhardt:2017aux}. 

We choose a gauge in null coordinates $u_{E},v_{E}$ and spatial coordinates $x^{i}$. Fixing $\mu$ at $u_{E}=0$ and $v_{E}=0$, we set $l_{E}^{a}=(\partial/\partial u_{E})^{a}$ and $k_{E}^{a}=(\partial/\partial v_{E})^{a}$. The gauge conditions are:
\begin{equation}
\begin{split}
g_{u_{E}v_{E}}&=-1\\g_{u_{E}u_{E}}&=g_{u_{E}i}=0\\g_{v_{E}v_{E}}|_{u_{E}=0}=g_{v_{E}v_{E},u_{E}}&|_{u_{E}=0}=g_{v_{E}i}|_{u_{E}=0}=0
\end{split}
\end{equation}
These gauge conditions imply that we are in the Gaussian null coordinates (GNC) \cite{Wall:2015raa, Wall:2024lbd, Hollands:2022fkn, Bhattacharyya:2021jhr, Visser:2024pwz, Hollands:2024vbe}
\begin{equation}
    ds^{2}_{E}=-2du_{E}dv_{E}-u_{E}^{2}\alpha dv_{E}^{2}-2u_{E}\omega_{i}dv_{E}dx^{i}+\gamma_{ij}dx^{i}dx^{j}.
\end{equation}
In this gauge, the constraint equations become \cite{Engelhardt:2018kcs, Hayward:2006ss, Hayward:2001ck, Hayward:2004fz, Gourgoulhon:2005ng, Cao:2010vj}
\begin{align}
    \theta_{u_{E},v_{E}}&=-\frac{1}{2}\mathcal{R}^{E}+\nabla^{E}\cdot\chi^{E}-\theta_{v_{E}}\theta_{u_{E}}+8\pi GT_{u_{E}v_{E}}+\chi_{E}^{2}\label{theta uv}\\ \theta_{v_{E},v_{E}}&=-\frac{1}{D-2}\theta_{v_{E}}^{2}-\varsigma_{v_{E}}^{2}-8\pi GT_{v_{E}v_{E}}\label{thetavv}\\ \chi^{E}_{i,v_{E}}&=-\theta_{v_{E}}\chi^{E}_{i}+(\frac{D-3}{D-2})\nabla_{i}\theta_{v_{E}}-(\nabla\cdot\varsigma_{v_{E}})_{i}+8\pi GT_{iv_{E}}\label{chi iv},
\end{align}
we will use $\theta_{u_{E}}$ and $\theta_{v_{E}}$ to emphasize our choice of gauge.

Constructing a stationary null hypersurface requires specifying initial data on $N_{-k_E}$ (the data on $N_{-l_E}$ being fixed in $O^{E}_W[\mu]$). As the procedure differs only slightly from that in \cite{Engelhardt:2018kcs}, we only provide a very simplified discussion. Specifically, we require that
\begin{align}
    \varsigma_{v_{E}}[N_{-k_{E}}]&=0\label{chi condition}\\ T^{E}_{v_{E}v_{E}}[N_{-k_{E}}]&=0\label{Tvv condition}\\T^{E}_{iv_{E}}[N_{-k_{E}}]&=0\label{Tiv condition}\\T^{E}_{u_{E}v_{E}}[N_{-k_{E}}]&=\text{const}\label{Tuv condition}.
\end{align}
Substituting \eqref{chi condition} and \eqref{Tvv condition} into \eqref{thetavv}, and noting that $\theta_{v_{E}}|_{v_{E}=0}=0$, implies $\theta_{v_{E}}=0$. Consequently, $N_{-k_{E}}$ is stationary, and $\mathcal{R}^{E}$ is also constant along $N_{-k_{E}}$. Substituting \eqref{Tiv condition} into \eqref{chi iv} reveals that $\chi^{E}_i$ is constant on $N_{-k_E}$. Together with \eqref{Tuv condition} and \eqref{theta uv}, it shows that $\theta_{u_{E},v_{E}}|_{u_{E}=0}=\text{constant}$. 

We now justify these requirements on the stress tensor. From \eqref{stress tensor in the Einstein frame}, the stress tensor in the Einstein frame splits into two parts. First, consider the contribution from $\phi$:
\begin{equation}
T^{\phi}_{ab}=\partial_{a}\phi\partial_{b}\phi-g_{ab}^{E}\Big(\frac{1}{2}g_{E}^{cd}\partial_{c}\phi\partial_{d}\phi+V(\phi)\Big).
\end{equation}
In the characteristic initial value problem, $\partial_{v_E}\phi$ is a free data on the null hypersurface $N_{-k_E}$ and can thus be set to zero. And in our gauge $g_{v_{E}v_{E}}|_{u_{E}=0}=g_{v_{E}i}|_{u_{E}=0}=0$, therefore
\begin{equation}
T^{\phi}_{v_{E}v_{E}}=T^{\phi}_{iv_{E}}=0.
\label{condition for Tphi}
\end{equation}
For the matter contribution $T^{m}_{ab}$ (here we consider complex scalar field for simplicity), as shown in \eqref{relation between Tm and T}
\begin{equation}
\begin{split}
    T^{m}_{v_{E}v_{E}}&=\frac{T_{v_{E}v_{E}}}{f'(R)}\\T^{m}_{iv_{E}}&=\frac{T_{iv_{E}}}{f'(R)}.
    \label{relation between Tm and T in section 3}
\end{split}
\end{equation}
Since null geodesics are conformally invariant under the Weyl transformation \eqref{Weyl transformation}, the null geodesic $k_{E}^{a}=(\partial_{v_{E}})^{a}$ remains null geodesic in the $f(R)$ frame but with a different affine parameter $v$, since $\partial_{v_{E}}\phi\propto\partial_{v_{E}}\text{log}f'(R)=0$, allowing us to set the parameter $v$ \cite{Wald:1984rg}
\begin{equation}
    \frac{dv_{E}}{dv}=1,
    \label{relation between two parameters}
\end{equation}
or simply $v_{E}=v$ (further details are provided in section \ref{Construction of Stationary Null Hypersurface For the Generalized Expansion}). Consequently, it is physically reasonable to set the corresponding stress tensor components $T_{iv_{E}} \text{ and }T_{v_{E}v_{E}}$ in the $f(R)$ frame to zero\footnote{Since for usual matter field, for example complex scalar $\Phi$, we can set $\partial_{v_{E}}\Phi=\partial_{v}\Phi=0$, this will lead $ T^{m}_{v_{E}v_{E}}=T^{m}_{iv_{E}}=0$. And this is also true for other usual matter field such as Maxwell Field \cite{Engelhardt:2018kcs}.}. Combining this with \eqref{condition for Tphi} and \eqref{relation between Tm and T in section 3} yields
\begin{equation}
    T^{E}_{v_{E}v_{E}}=T^{E}_{iv_{E}}=0
\end{equation}
is a very reasonable setting. The constancy of $T^{E}_{u_{E}v_{E}}$ can be derived by the Bianchi identity and the Gauss Law on the null hypersurface. We expect that a similar prescription exists for other reasonable matter fields to satisfy \eqref{Tvv condition} \eqref{Tiv condition} and \eqref{Tuv condition}. Assuming so, it is always possible to construct a stationary null hypersurface $N_{-k_{E}}$ satisfying the constraint equations.

To ensure continuity across the gluing surface, we must verify the junction conditions. The junction condition for $\theta_{k_{E}}$ is already satisfied, as $N_{-k_{E}}$ is a stationary null hypersurface where $\theta_{k_{E}}$ vanishes, matching its value on $\mu$. Regarding the remaining junction conditions, our gauge fixes the transverse metric $g_{ij}^E$, the twist $\chi_i$, and $\theta_{u_E, v_E}[N_{-k_E}]$ only up to functions of $x^i$. Even after fixing $\theta_{u_E, v_E}[N_{-k_E}]$, we retain the freedom to specify $\theta_{u_E}$, as well as the fields $\phi$ and $\Phi$. We are therefore free to choose all the quantities to be continuous across $\mu$
\begin{align}
    g_{ij}^{E}[N_{-k_{E}}]&=g^{E}_{ij}[\mu]\\\chi_{i}[N_{-k_{E}}]&=\chi_{i}[\mu]\\\theta_{u_{E},v_{E}}[N_{-k_{E}}]&=\theta_{u_{E},v_{E}}[\mu]\\\theta_{u_{E}}[N_{-k_{E}}]|_{v_{E}=0}&=\theta_{u_{E}}[\mu]\\\phi[N_{-k_{E}}]&=\phi[\mu]\label{phi condition}\\\Phi[N_{-k_{E}}]&=\Phi[\mu]\label{Phi condition},
\end{align}
and the last two conditions \eqref{phi condition} and \eqref{Phi condition} guarantee that $T^{E}_{u_{E}v_{E}}[N_{-k_{E}}]=T^{E}_{u_{E}v_{E}}[\mu]$. We conclude that this choice satisfies the junction conditions and ensures field continuity in the Einstein frame.

On $N_{-k_{E}}$, $\theta_{u_E, v_E}^E$ is constant on slices of constant $v$. From the minimar condition on $\mu$, it follows that $\theta_{u_E, v_E}^E|_{v_E} < 0$. In addition, we know that $\theta^{E}_{u_{E}}[\mu]<0$, this will guarantee that there will be a surface $X$ on $N_{-k}$ which $\theta^{E}_{u_{E}}=\theta^{E}_{v_{E}}=0$. This surface $X$ need not be a slice of constant $v_{E}$; generally, $X$ is defined by $v_E = h(x^i)$. Our conditions ensure the existence of $h(x^i)$. Then we apply the CPT reflection in the Einstein frame across $X$ that takes $v_{E}\to-v_{E}$, $u_{E}\to-u_{E}$, $x^i\to x^{i}$. It is easy to show that all quantities that are odd under the CPT reflection vanish on $X$ in our construction. Therefore, the junction conditions are automatically satisfied across $X$. Given the initial data constructed on the Cauchy slice $\Sigma=N_{-l}[\mu]\cup N_{-k}[\mu]\tilde{N}_{-l}[\tilde{\mu}]$, the entire spacetime $(\mathcal{M}^{\{\alpha\}},g^{E})$ is well-defined.

Finally, assuming the NEC holds for matter in the $f(R)$ frame, the NCC is satisfied in the Einstein frame. This will give us the focusing theorem in the Einstein frame
\begin{equation}
    \theta_{v_{E},v_{E}}=-\frac{1}{D-2}\theta_{v_{E}}^{2}-\varsigma_{v_{E}}^{2}-8\pi GT_{v_{E}v_{E}}\leq0.
\end{equation}
This implies that the constructed extremal surface $X$ is the HRT surface in $(\mathcal{M}^{\{\alpha\}}, g^E)$; specifically, any other extremal surface $X'$ must have an area greater than that of $X$ \cite{Engelhardt:2017aux, Engelhardt:2018kcs}. Then the von
Neumann entropy $S_{\text{vN}}(\rho^{E}_{B})$ of $(\mathcal{M}^{\{\alpha\}},g^{E})$
\begin{equation}
    S_{\text{vN}}(\rho^{E}_{B})=\frac{\text{Area}_{E}[\mu]}{4G}.
\end{equation}
Thus, in the Einstein frame, we have
\begin{equation}
    S^{\text{(outer)}}_{E}[\mu]=\frac{\text{Area}_{E}[\mu]}{4G}.
\end{equation}

\section{Relation Between the Outer Entropy in the Einstein Frame and the \texorpdfstring{$f(R)$}{f(R)} Frame}
\label{Relation Between the Outer Entropy in Einstein Frame and f(R) frame}

In this section, we establish a correspondence between the von Neumann entropies in the Einstein and $f(R)$ frames. More precisely, we demonstrate that at a given boundary time $t$,
\begin{equation}
    S_{\text{vN}}\big(\rho^{E}_{A}(t)\big)=S_{\text{vN}}\big(\rho^{f}_{A}(t)\big),
    \label{von Neumann entropy correspondence}
\end{equation}
where $A$ is a general subregion of the boundary Cauchy slice $\Sigma_{t}$, and $\rho_{A}^{E}$ and $\rho_{A}^{f}$ are the reduced density matrices of the subregion $A$ in the Einstein and $f(R)$ frames, respectively. Furthermore, their corresponding bulk duals are related by the transformations described in section \ref{section Einstein Frame}. We will show \eqref{von Neumann entropy correspondence} by constructing the correspondence between the gravity dual of a Schwinger-Keldysh contour \cite{Dong:2016hjy} in both frames. We first provide a brief review of the Schwinger-Keldysh contour on the boundary and its gravity dual, based on Xi Dong, Aitor Lewkowycz and Mukund Rangamani's construction \cite{Dong:2016hjy}. 

Next, we construct the correspondence between the gravity dual of the Schwinger-Keldysh contour, and show the correspondence of von Neumann entropy between the Einstein frame and the $f(R)$ frame. Moreover, we consider the time reflection symmetric case and derive the RT formula of $f(R)$ gravity. And we find that our results agree with the results of Xi Dong \cite{Dong:2013qoa} as a correctness check.

Finally, we show that the outer entropy of a codimension-2 surface $\sigma$ is identical in both frames, if the outer wedge $O_{W}^{E}[\sigma]$ and $O_{W}^{f}[\sigma]$ are related by the transformation between the Einstein frame and the $f(R)$ frame. Then we show that the outer entropy of the generalized marginally trapped surface is just the Wald entropy associated with it.

\subsection{The Schwinger-Keldysh Contour and its Gravity Dual}

\begin{figure}[H]
    \centering

\tikzset{every picture/.style={line width=0.75pt}} 

\begin{tikzpicture}[x=0.75pt,y=0.75pt,yscale=-1,xscale=1]

\draw    (89.6,70.33) -- (244.22,70) ;
\draw    (89.6,70.33) .. controls (82.12,100.67) and (75.23,142.67) .. (28.67,160) ;
\draw    (244.22,70) .. controls (236.75,101.33) and (209.73,158.67) .. (183.87,160) ;
\draw    (28.67,160) -- (183.87,160) ;
\draw  [dash pattern={on 4.5pt off 4.5pt}]  (89.6,70.33) .. controls (91.9,101.33) and (98.79,140) .. (132.13,140.67) ;
\draw    (244.22,70) .. controls (248.25,107.33) and (251.69,138.67) .. (287.33,140.67) ;
\draw  [dash pattern={on 4.5pt off 4.5pt}]  (132.13,140.67) -- (210.31,140.67) ;
\draw    (210.31,140.67) -- (287.33,140.67) ;
\draw    (176.97,150.67) .. controls (188.18,141.57) and (192.83,135) .. (199.98,114.29) ;
\draw [shift={(200.54,112.67)}, rotate = 108.76] [color={rgb, 255:red, 0; green, 0; blue, 0 }  ][line width=0.75]    (10.93,-3.29) .. controls (6.95,-1.4) and (3.31,-0.3) .. (0,0) .. controls (3.31,0.3) and (6.95,1.4) .. (10.93,3.29)   ;
\draw  [dash pattern={on 4.5pt off 4.5pt}]  (110.29,83.33) .. controls (113.06,108.42) and (111.56,106.82) .. (117.13,122.25) ;
\draw [shift={(117.76,124)}, rotate = 249.96] [color={rgb, 255:red, 0; green, 0; blue, 0 }  ][line width=0.75]    (10.93,-3.29) .. controls (6.95,-1.4) and (3.31,-0.3) .. (0,0) .. controls (3.31,0.3) and (6.95,1.4) .. (10.93,3.29)   ;
\draw    (349.05,70.85) .. controls (341.8,101.01) and (335.12,142.77) .. (290,160) ;
\draw    (489.98,80.79) .. controls (482.73,111.95) and (465.47,158.67) .. (440.4,160) ;
\draw    (290,160) -- (440.4,160) ;
\draw  [dash pattern={on 4.5pt off 4.5pt}]  (349.05,70.85) .. controls (351.27,101.67) and (357.96,140.12) .. (390.27,140.78) ;
\draw    (494.43,66.21) .. controls (498.33,103.33) and (506.13,138.79) .. (540.67,140.78) ;
\draw  [dash pattern={on 4.5pt off 4.5pt}]  (390.27,140.78) -- (466.02,140.78) ;
\draw    (466.02,140.78) -- (540.67,140.78) ;
\draw    (433.72,150.72) .. controls (444.58,141.67) and (449.09,135.15) .. (456.02,114.55) ;
\draw [shift={(456.55,112.94)}, rotate = 108.32] [color={rgb, 255:red, 0; green, 0; blue, 0 }  ][line width=0.75]    (10.93,-3.29) .. controls (6.95,-1.4) and (3.31,-0.3) .. (0,0) .. controls (3.31,0.3) and (6.95,1.4) .. (10.93,3.29)   ;
\draw  [dash pattern={on 4.5pt off 4.5pt}]  (369.1,83.78) .. controls (371.79,108.72) and (370.32,107.13) .. (375.73,122.47) ;
\draw [shift={(376.34,124.21)}, rotate = 250.43] [color={rgb, 255:red, 0; green, 0; blue, 0 }  ][line width=0.75]    (10.93,-3.29) .. controls (6.95,-1.4) and (3.31,-0.3) .. (0,0) .. controls (3.31,0.3) and (6.95,1.4) .. (10.93,3.29)   ;
\draw    (349.05,70.85) -- (433.16,70.85) ;
\draw    (433.16,70.85) -- (489.98,80.79) ;
\draw    (433.16,70.85) -- (494.43,66.21) ;

\draw (122.97,94.67) node   [align=left] {$\displaystyle t$};
\draw (187.93,118) node   [align=left] {$\displaystyle t$};
\draw (381.39,95.04) node   [align=left] {$\displaystyle t$};
\draw (444.33,118.24) node   [align=left] {$\displaystyle t$};
\draw (68.23,70.85) node    {$\Sigma _{t}$};
\draw (328.34,70.04) node    {$\Sigma _{t}$};
\draw (476.52,55.47) node    {$A$};
\draw (373.21,55.81) node    {$A^{c}$};
\draw (428.91,55.47) node    {$\partial A$};
\draw (384.33,172.4) node [anchor=north west][inner sep=0.75pt]    {(b)};
\draw (117.33,172.73) node [anchor=north west][inner sep=0.75pt]    {(a)};

\end{tikzpicture}
\caption{(a) is the Schwinger-Keldysh construction for $\text{Tr}\rho(t)$. (b) is the Schwinger-Keldysh construction for the reduced density matrix $\rho_{A}(t)$.}
\label{Figure 1}
\end{figure}
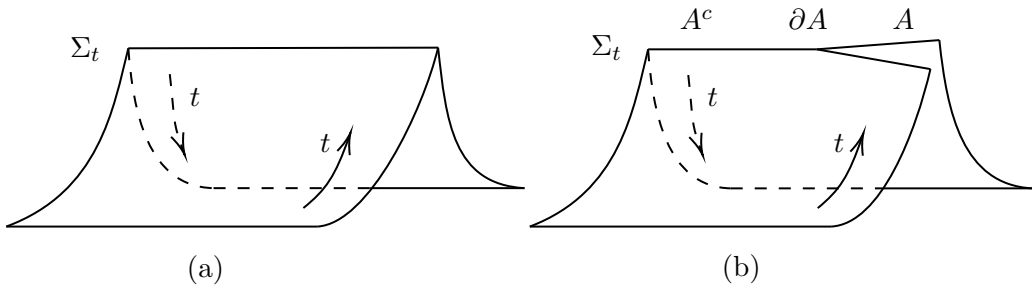

In this section, we will give a short review of Schwinger-Keldysh contour and its gravity dual. Since the Cauchy slice $\Sigma_t$ generally lacks time-reflection symmetry, we cannot directly apply the path integral from the far past to far future. Instead, we must employ the Keldysh contour \cite{Keldysh:1964ud, Dong:2016hjy}, where the expression for $\rho(t)$ is given by
\begin{equation}
    \rho(t)=|C_{t}\rangle \langle C_{t}|=\int[D\Phi]e^{iS_{\uparrow}[\Phi]-iS_{\downarrow}[\Phi]},
\end{equation}
where $|C_t\rangle$ is the boundary state at time $t$; the up-arrow indicates forward time evolution, and the down-arrow indicates backward time evolution. This can be well appreciated in the picture, see the figure \ref{Figure 1} (a).

After we obtain the path integral representation of $\rho(t)$, we should use this to compute entanglement entropy of subregion $A$ with $\partial A$ as the entangling surface. To calculate the reduced density matrix $\rho_A(t)$, we introduce a cut along $A$ and use $t^\pm$ to denote the evolution of the two legs. And we fix some boundary conditions along $A$: $\Phi(t=t^{+})|_{A}=\Phi^{+}$ for the forward part and $\Phi(t=t^{-})|_{A}=\Phi^{-}$ for the backward part. We will obtain the matrix element $(\rho_{A})_{+-}$ \cite{Dong:2016hjy}, the diagram representation as shown in figure \ref{Figure 1} (b).

There is a useful Rindler coordinate chart in the vicinity of $\partial A$. Now we consider a simple example, the boundary geometry is flat spacetime $B=\mathbb{R}^{1,D-1}$. And we choose the Cartesian coordinates $(t,x,y^{i})$. In these coordinates, $\Sigma_{t}$ is the $t=0$ Cauchy slice and $A$ is located at $x>0$. Note that $D^-[A]$ corresponds to the past half of the right Rindler wedge in flat spacetime. Therefore, now we consider the Rindler chart
\begin{equation}
    ds^{2}=-dt^{2}+dx^{2}+dy^{i}dy^{i}=-r^{2}d\tau^{2}+dr^{2}+dy^{i}dy^{i}.
\end{equation}
The advantage of choosing this coordinate chart is that we can include all spacetime regions, by allowing $\tau$ to be complex with a discrete imaginary part. To be precise, let $\tau=\tau_{A}+\frac{m\pi}{2}i$ where $\tau_{A}$ is real and $m=\text{0,1,2,3,4}$. The regions corresponding to $m=0$ ($\tau_A<0$, representing the forward part of $D^-[A]$) and $m=4$ ($\tau_A>0$, representing the backward part of $D^-[A]$) are the most crucial. To compute $\text{Tr}(\rho_A)$, we glue the $m=0$ domain to the $m=4$ domain along $A$. Therefore, we identify $\tau\sim\tau+2\pi i$ \cite{Dong:2016hjy}.

Once we obtain the reduced density matrix, we can calculate the entanglement entropy by replica trick
\begin{equation}
    S_{\text{vN}}(\rho_{A})=-\text{Tr}_{A}(\rho_{A}\text{log}\rho_{A})=\lim_{q\to1}\frac{1}{1-q}\text{log}\ \text{Tr}\rho_{A}^{q}=\lim_{q\to1}S_{A}^{(q)},
\end{equation}
here $S^{(q)}_{A}$ is the R\'enyi entropy. Computing the Rényi entropy thus requires evaluating $\rho_A^q$. In the gluing construction, we have $q$ different coordinates and the gluing condition is $\tau_{J}=\tau_{J-1}+2\pi i$ along $A$. However, it is more convenient to introduce a single coordinate $\tau$ with $4q+1$ discrete imaginary parts, that is, $\tau=\tau_{A}+\frac{m\pi}{2}i$ here now $m=0,1,2\cdots,4q$. And in this coordinate, the gluing condition is $\tau\sim\tau+2\pi iq$ \cite{Dong:2016hjy}.

There is a $\mathbb{Z}_{q}$ symmetry originating from the invariance under the exchange of replicas of $\rho_{A}$. Obviously, $\partial A$ is the fixed surface under the $\mathbb{Z}_{q}$ symmetry. We will assume that this symmetry is unbroken. And this symmetry will ensure that every copy of the Schwinger-Keldysh contour is the same. Now we need to construct the gravity dual of the Schwinger-Keldysh contour. We assume that the bulk fields satisfy both the gluing condition ($\tau \sim \tau + 2\pi i q$) and the $\mathbb{Z}_q$ replica symmetry.

\begin{figure}[h]
    \centering

\tikzset{every picture/.style={line width=0.75pt}} 

\begin{tikzpicture}[x=0.75pt,y=0.75pt,yscale=-1,xscale=1]

\draw    (51.33,124.67) -- (51.33,207.33) ;
\draw    (170.67,124.67) -- (170.67,207.33) ;
\draw  [draw opacity=0][dash pattern={on 4.5pt off 4.5pt}] (51.55,124.29) .. controls (53.96,113.99) and (79.69,105.88) .. (111.07,105.88) .. controls (142.82,105.88) and (168.79,114.19) .. (170.67,124.67) -- (111.07,125.88) -- cycle ; \draw  [dash pattern={on 4.5pt off 4.5pt}] (51.55,124.29) .. controls (53.96,113.99) and (79.69,105.88) .. (111.07,105.88) .. controls (142.82,105.88) and (168.79,114.19) .. (170.67,124.67) ;  
\draw  [draw opacity=0] (170.3,123.87) .. controls (170.52,124.57) and (170.64,125.28) .. (170.64,126) .. controls (170.64,137.05) and (143.92,146) .. (110.96,146) .. controls (78.81,146) and (52.6,137.48) .. (51.33,126.81) -- (110.96,126) -- cycle ; \draw   (170.3,123.87) .. controls (170.52,124.57) and (170.64,125.28) .. (170.64,126) .. controls (170.64,137.05) and (143.92,146) .. (110.96,146) .. controls (78.81,146) and (52.6,137.48) .. (51.33,126.81) ;  
\draw  [draw opacity=0][dash pattern={on 4.5pt off 4.5pt}] (51.44,205.08) .. controls (53.86,194.77) and (79.58,186.67) .. (110.96,186.67) .. controls (142.71,186.67) and (168.68,194.97) .. (170.56,205.45) -- (110.96,206.67) -- cycle ; \draw  [dash pattern={on 4.5pt off 4.5pt}] (51.44,205.08) .. controls (53.86,194.77) and (79.58,186.67) .. (110.96,186.67) .. controls (142.71,186.67) and (168.68,194.97) .. (170.56,205.45) ;  
\draw  [draw opacity=0] (170.3,204.53) .. controls (170.52,205.23) and (170.64,205.95) .. (170.64,206.67) .. controls (170.64,217.71) and (143.92,226.67) .. (110.96,226.67) .. controls (78.81,226.67) and (52.6,218.15) .. (51.33,207.48) -- (110.96,206.67) -- cycle ; \draw   (170.3,204.53) .. controls (170.52,205.23) and (170.64,205.95) .. (170.64,206.67) .. controls (170.64,217.71) and (143.92,226.67) .. (110.96,226.67) .. controls (78.81,226.67) and (52.6,218.15) .. (51.33,207.48) ;  
\draw    (102,146) .. controls (76,54) and (102.67,76.67) .. (113.33,105.33) ;
\draw    (113.33,105.33) .. controls (134,56) and (134.67,77.33) .. (102,146) ;
\draw    (51.44,124.41) .. controls (56.67,106) and (73.33,86) .. (92.77,81.08) ;
\draw    (128,78.67) .. controls (147.33,80) and (168.67,102) .. (170.67,124.67) ;
\draw    (58.53,93.86) .. controls (71.53,33.85) and (188.57,30.96) .. (208.75,88.89) ;
\draw [shift={(209.33,90.67)}, rotate = 252.9] [color={rgb, 255:red, 0; green, 0; blue, 0 }  ][line width=0.75]    (10.93,-4.9) .. controls (6.95,-2.3) and (3.31,-0.67) .. (0,0) .. controls (3.31,0.67) and (6.95,2.3) .. (10.93,4.9)   ;
\draw [shift={(58,96.67)}, rotate = 278.97] [color={rgb, 255:red, 0; green, 0; blue, 0 }  ][line width=0.75]    (10.93,-4.9) .. controls (6.95,-2.3) and (3.31,-0.67) .. (0,0) .. controls (3.31,0.67) and (6.95,2.3) .. (10.93,4.9)   ;
\draw    (194.67,124.67) -- (194.67,207.33) ;
\draw    (314,124.67) -- (314,207.33) ;
\draw  [draw opacity=0][dash pattern={on 4.5pt off 4.5pt}] (194.77,124.41) .. controls (197.19,114.11) and (222.91,106) .. (254.29,106) .. controls (286.05,106) and (312.01,114.3) .. (313.89,124.78) -- (254.29,126) -- cycle ; \draw  [dash pattern={on 4.5pt off 4.5pt}] (194.77,124.41) .. controls (197.19,114.11) and (222.91,106) .. (254.29,106) .. controls (286.05,106) and (312.01,114.3) .. (313.89,124.78) ;  
\draw  [draw opacity=0] (313.63,123.87) .. controls (313.86,124.57) and (313.97,125.28) .. (313.97,126) .. controls (313.97,137.05) and (287.25,146) .. (254.29,146) .. controls (222.14,146) and (195.93,137.48) .. (194.66,126.81) -- (254.29,126) -- cycle ; \draw   (313.63,123.87) .. controls (313.86,124.57) and (313.97,125.28) .. (313.97,126) .. controls (313.97,137.05) and (287.25,146) .. (254.29,146) .. controls (222.14,146) and (195.93,137.48) .. (194.66,126.81) ;  
\draw  [draw opacity=0][dash pattern={on 4.5pt off 4.5pt}] (194.77,205.08) .. controls (197.19,194.77) and (222.91,186.67) .. (254.29,186.67) .. controls (286.05,186.67) and (312.01,194.97) .. (313.89,205.45) -- (254.29,206.67) -- cycle ; \draw  [dash pattern={on 4.5pt off 4.5pt}] (194.77,205.08) .. controls (197.19,194.77) and (222.91,186.67) .. (254.29,186.67) .. controls (286.05,186.67) and (312.01,194.97) .. (313.89,205.45) ;  
\draw  [draw opacity=0] (313.63,204.53) .. controls (313.86,205.23) and (313.97,205.95) .. (313.97,206.67) .. controls (313.97,217.71) and (287.25,226.67) .. (254.29,226.67) .. controls (222.14,226.67) and (195.93,218.15) .. (194.66,207.48) -- (254.29,206.67) -- cycle ; \draw   (313.63,204.53) .. controls (313.86,205.23) and (313.97,205.95) .. (313.97,206.67) .. controls (313.97,217.71) and (287.25,226.67) .. (254.29,226.67) .. controls (222.14,226.67) and (195.93,218.15) .. (194.66,207.48) ;  
\draw    (245.33,146) .. controls (219.33,54) and (246,76.67) .. (256.67,105.33) ;
\draw    (256.67,105.33) .. controls (277.33,56) and (278,77.33) .. (245.33,146) ;
\draw    (194.77,124.41) .. controls (200,106) and (216.67,86) .. (236.1,81.08) ;
\draw    (271.33,78.67) .. controls (290.67,80) and (312,102) .. (314,124.67) ;

\draw (155.39,149.2) node    {$A$};
\draw (66.13,149) node    {$A^{c}$};
\draw (144.64,123.52) node    {$\mathcal{R}_{A}$};
\draw (40.89,112.25) node    {$\tilde{\Sigma }_{t}$};
\draw (298.72,149.2) node    {$A$};
\draw (209.47,149) node    {$A^{c}$};
\draw (287.97,123.52) node    {$\mathcal{R}_{A}$};
\draw (184.22,112.25) node    {$\tilde{\Sigma }_{t}$};
\draw (132.42,61.83) node   [align=left] {identify};
\draw (81.03,125.08) node    {$\mathcal{R}_{A^{c}}$};
\draw (220.36,123.75) node    {$\mathcal{R}_{A^{c}}$};

\end{tikzpicture}
\caption{This diagram is about the path integral construction of reduced density matrix $\rho_{A}(t)$ in the bulk. The gravity correspondence of tracing the degree of freedom of $A^{c}$ is gluing the geometry across $\mathcal{R}_{A^{c}}$.}
\label{Gravity Dual of Schwinger-Keldysh}
\end{figure}
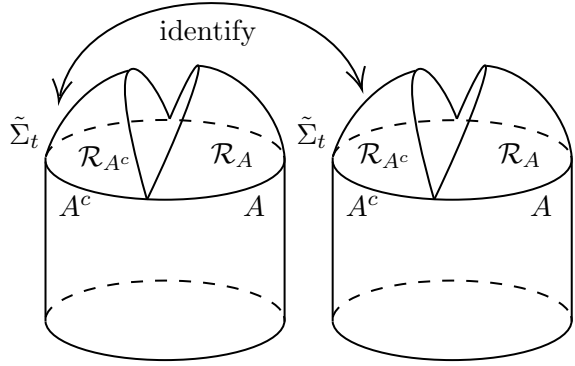

From the preceding discussion, it is evident that the gravity dual cannot be constructed in a Euclidean manifold in the absence of time-reflection symmetry. Instead, we try to construct the gravity dual of the Schwinger-Keldysh contour in Lorentzian spacetime. It was shown that in the bulk there is a Cauchy slice $\tilde{\Sigma}_t$ in the Wheeler-DeWitt patch of $\Sigma_t$ as the bulk dual of $\Sigma_t$ ($\partial\tilde{\Sigma}_{t}=\Sigma_t$). And the bulk evolution only contains the past of the $\tilde{\Sigma}_{t}$, i.e., in $\tilde{J}^{-}[\tilde{\Sigma}_{t}]$. In other words the initial conditions are evolved forward from $t=-\infty$ up to $\tilde{\Sigma}_{t}$ and then we evolve back to construct the bulk Schwinger-Keldysh contour \cite{Dong:2016hjy}. And there should exist a co-dimension 2 surface $e$ as the bulk dual of the $\partial A$, and it is natural to have $\partial e=\partial A$. These can be clearly represented by figure \ref{Gravity Dual of Schwinger-Keldysh}.

Now we try to show how to construct the bulk dual of the reduced density matrix $\rho_{A}(t)$, we first consider the case $q=1$. We start with the fact that we can divide the Cauchy slice $\tilde{\Sigma}_{t}$ into two regions $\tilde{\Sigma}_{t}=\mathcal{R}_{A^{c}}\cup\mathcal{R}_{A}$, with an intersection co-dimension 2 surface $e$. To understand how we extend the boundary coordinates into the bulk, we foliate the causal development of $\mathcal{R}_{A}$ in Rindler-like coordinates. In general, in the vicinity of any co-dimension 2 surface such as $e$, we can naturally write the metric as
\begin{equation}
    ds^{2}=-r^{2}d\tau^{2}+dr^{2}+(\gamma_{ij}+re^{\pm\tau}K^{\pm}_{ij}+\cdots)dy^{i}dy^{j}.
\end{equation}
For a Rindler like observer, there will be four horizons, just like the previous construction, a horizon crossing can be understood as $\tau\to\tau+\frac{\pi}{2}i$ (with $r\to i^{-1}r$) \cite{Dong:2016hjy}. It captures the local geometry in a neighborhood of $e$ efficiently. When we calculate $\text{Tr}_{A}\rho_{A}$, we will return to the starting point after crossing four horizons. This will be encoded in the gluing condition $\tau\sim\tau+2\pi i$.

Now we consider the case where $q\neq1$. Defining the corresponding bulk geometry as $\mathcal{M}_{q}$, it is natural to consider a geometry that has the same gluing condition ($\tau\sim\tau+2\pi iq$) and replica $\mathbb{Z}_{q}$ symmetry. And the surface $e_{q}$ the bulk correspondence of $\partial A$, should invariant under replica $\mathbb{Z}_{q}$ symmetry. We also assume that the partition function of boundary should be computed in the saddle point approximation of the bulk geometry $\mathcal{M}_{q}$ which satisfies all the symmetries and bulk EOM without any singularity
\begin{equation}
\text{Tr}_{A}\rho_{A}^{q}=I_{q}.
\end{equation}
Here $I_{q}$ is the corresponding on-shell action of the bulk physics, which consists of three parts
\begin{equation}
I_{q}=I[\mathcal{M}_{q}]+I[B_{q}]+I[\tilde{\Sigma}_{t}].
\end{equation}
Due to the $\mathbb{Z}_q$ symmetry around $e_q$, we need only focus on a single copy of the replica geometry $\hat{\mathcal{M}}_{q}=\mathcal{M}_{q}/\mathbb{Z}_{q}$
\begin{equation}
    I_{q}=qI_{1}=q(I[\hat{\mathcal{M}_{q}}]+I[\text{boundaries}]).
\end{equation}
The one copy geometry $\hat{\mathcal{M}}_{q}$ has $\mathbb{Z}_{q}$ symmetry fixed surface $e_{q}$ with conical singularity \cite{Gutperle:2002ai}. And the result of von Neumann entropy from the bulk side is
\begin{equation}
    S_{\text{vN}}\big(\rho_{A}(t)\big)=\lim_{q\to 1}\partial_{q}I_{1}.
\end{equation}

To calculate this derivative $\partial_{q}$, we should regard this as a variation of the bulk solution, since the conical structure will only affect the geometry in the vicinity of $e_{q}$. And the boundary geometry remains unchanged, $\partial_{q}g_{q}|_{\partial\mathcal{M}_{q}\text{or}\tilde{\Sigma}_{t}}=0$. Therefore, the boundary terms will not contribute to the von Neumann entropy. We only need to consider the bulk term $I[\hat{\mathcal{M}_{q}}]$. Under the standard variation process \cite{Dong:2016hjy}
\begin{equation}
    \partial_{q}I[\hat{\mathcal{M}_{q}}]=\int_{\mathcal{M}_{q}}\big(\text{EOM}\cdot\partial_{q}g_{q}+d\Theta(g_{q},\partial_{q}g_{q})\big)=\int_{e_{q}(\epsilon)}\Theta(g_{q},\partial_{q}g_{q}),
\end{equation}
where we have chosen to regulate the result by blowing up the singular locus to a tubular neighbourhood. We will obtain the answer when $\epsilon\to 0$.

\subsection{Correspondence of von Neumann Entropy}

In this section, we will use the results in the last section to prove the \eqref{von Neumann entropy correspondence}. We now aim to prove that if the geometries are related by the Weyl transformation between the Einstein and $f(R)$ frames, their corresponding Rényi entropies—or more precisely, $I[\mathcal{M}_q]$—are identical:
\begin{equation}
    I^{E}[\mathcal{M}_{q}]=I^{f}[\mathcal{M}_{q}].
    \label{correspondence between Renyi entropy}
\end{equation}

We first construct a replica geometry $(g_{q}^{E},\phi,\text{matter fields},\mathcal{M}_{q})$ in the Einstein frame, exhibiting gluing condition and $\mathbb{Z}_q$ replica symmetry with respect to the codimension-2 surface $e_q$. And surface $e_{q}$ give a partition in Cauchy slice $\tilde{\Sigma}_{t}$ (is located in the Wheeler-DeWitt patch of $\Sigma_{t}$) whose boundary is $\Sigma_{t}$. This geometry satisfies the Einstein equation without any singularity. When $q=1$ this geometry will return to the geometry corresponding to boundary density matrix $\rho^{E}(t)$.

Now we move into the $f(R)$ frame, from the inverse transformations \eqref{inverse transformation 1} \eqref{inverse transformation 2} and \eqref{inverse transformation 3}, it is easy to show that:
\begin{enumerate}
    \item If in the Einstein frame $g_{q}^{E}$ $\phi$ and other matter fields satisfy the gluing condition and replica $\mathbb{Z}_{q}$ symmetry, after we move into the $f(R)$ frame, the geometry ($g_{q}$, matter fields, $\mathcal{M}_{q}$) will also satisfy the gluing condition and replica $\mathbb{Z}_{q}$ symmetry.
    \item Since we do the Weyl transformation, the Wheeler-DeWitt patch of $\Sigma_{t}$ remains unchanged. Therefore, in the $f(R)$ frame $\tilde{\Sigma}_{t}$ is still located in the Wheeler-DeWitt patch of $\Sigma_{t}$.
    \item The surface $e_q$ remains fixed under the $\mathbb{Z}_q$ symmetry in the $f(R)$ frame. Furthermore, the entire spacetime satisfies the $f(R)$ equations of motion and is free of singularities.
    \item When $q=1$ the geometry will return to the geometry corresponding to boundary density matrix $\rho^{f}(t)$, and this geometry related to the geometry corresponding to $\rho^{E}(t)$ by the Weyl transformation between $f(R)$ and the Einstein frame.
\end{enumerate}
These will make the geometry ($g_{q}$, matter fields, $\mathcal{M}_{q}$) is the replica geometry of boundary density matrix $\rho^{f}(t)$. Conversely, the reverse implication also holds. If in the $f(R)$ frame we construct a replica geometry ($g_{q}$, matter fields, $\mathcal{M}_{q}$), after we move into the Einstein frame ($g^{E}_{q}$, $\phi$, matter fields, $\mathcal{M}_{q}$) remains a well defined replica geometry of boundary density matrix $\rho^{E}(t)$.

From the discussions in section \ref{section Einstein Frame}, it is evident that there exists a correspondence between the bulk actions
\begin{equation}
    I^{E}[\mathcal{M}_{q}]=I^{f}[\mathcal{M}_{q}],
    \label{relation between bulk action}
\end{equation}
if the geometries correspondent to $\rho^{E}(t)$ and $\rho^{f}(t)$ are related by the Weyl transformation between the $f(R)$ frame and the Einstein frame. Since $\mathbb{Z}_{q}$ is satisfied in both frames we can write
\begin{equation}
    qI^{E}[\hat{\mathcal{M}}_{q}]=qI^{f}[\hat{\mathcal{M}}_{q}].
    \label{relation between replica action}
\end{equation}
Following this, we perform an analytic continuation in $q$. Finally, the direct consequence of \eqref{relation between bulk action} and \eqref{relation between replica action} is
\begin{equation}
     S_{\text{vN}}\big(\rho^{E}_{A}(t)\big)=S_{\text{vN}}\big(\rho^{f}_{A}(t)\big).
\end{equation}

Now we apply our construction to a special case, when the spacetime satisfies the time reflection symmetry $t\to -t$ and consider the $t=0$ surface. Now we move into the Einstein frame, and from the above construction the time reflection symmetry remains unchanged. Since the time reflection symmetry, the extrinsic curvature of $t$ direction of $e_{q}$, $K^{E,t}$ will automatically be zero. We also require that the spacetime $\mathcal{M}_q$ contains no singularities. We will find that the extrinsic curvature of $x$ direction of $e_{q}$, $K^{E,x}$ will be zero \cite{Dong:2016hjy}. Here, $x$ and $t$ are bulk coordinates correspondent to boundary Cartesian coordinates. The constraint is equivalent to the minimal surface condition \cite{Ryu:2006bv} in the Einstein frame and the minimal Wald entropy in the $f(R)$ frame. And the von Neumann entropy is
\begin{equation}
    S_{\text{vN}}(\rho_{A}^{E})=S_{\text{vN}}(\rho^{f}_{A})=\min_{e_{q}\in \tilde{\Sigma}_{0}}\frac{\text{Area}_{E}[e_{q}]}{4G}=\min_{e_{q}\in \tilde{\Sigma}_{0}}\frac{\int_{e_{q}}f'(R)}{4G},
\end{equation}
this agrees with the result in \cite{Dong:2013qoa}. This is a self consistency check of our theory. 

\subsection{Relation Between Outer Entropy in Two Frames}

Now we consider the outer entropy of a co-dimension 2 surface $\sigma$ in the $f(R)$ frame. The outer entropy associated with a surface $\sigma$ homologous to $B$ is defined by maximizing the von Neumann entropy over all possible inner wedge data $\{\alpha\}$, while keeping the outer wedge $O_W^f[\sigma]$ fixed in the $f(R)$ frame:
\begin{equation}
    S^{\text{(outer)}}_{f}[\sigma]=\max_{\{\alpha\}}[S_{\text{vN}}(\rho^{f\{\alpha\}}_{B})].
\end{equation}
Here fixing the outer wedge $O^{f}_{W}[\sigma]$ means that we fix the geometry $g$ and matter fields in $O^{f}_{W}[\sigma]$. Transforming to the Einstein frame, we note that the Weyl transformation leaves the outer wedge invariant:
\begin{equation}
    O^{f}_{W}[\sigma]=O^{E}_{W}[\sigma].
\end{equation}
From the discussions in section \ref{section Einstein Frame}, when we fix $g$ and matter fields in $O^{f}_{W}[\sigma]$, $g^{E}$, $\phi$ and matter fields in the outer wedge $O^{E}_{W}[\sigma]$ in the Einstein frame will be fixed at the same time. Therefore, when we change the inner wedge data and fix the outer wedge in the $f(R)$ frame, in the Einstein frame the outer wedge $O^{E}_{W}[\sigma]$ is automatically fixed. 

From Section 2, for every asymptotically AdS solution in the $f(R)$ frame with a fixed outer wedge $O_W^f[\sigma]$ (satisfying the EOM and free of singularities), there exists a corresponding asymptotically AdS solution in the Einstein frame with a fixed outer wedge $O_W^E[\sigma]$ and identical on-shell action (satisfying the Einstein-frame EOM and free of singularities). As demonstrated in the previous subsection, their von Neumann entropies are identical:
\begin{equation}
    S_{\text{vN}}(\rho_{B}^{f\{\alpha\}})=S_{\text{vN}}(\rho_{B}^{E\{\alpha\}})=\frac{\text{Area}_{E}[X]}{4G}=\frac{\int_{X}f'(R)}{4G},
\end{equation}
where $X$ is the HRT surface ($\theta^{E}_{k_{E}}=\theta^{E}_{l_{E}}=0$) in the Einstein frame and the generalized extremal surface ($\Theta_{k}=\Theta_{l}=0$) in the $f(R)$ frame. Therefore, if we are maximizing von Neumann entropy in the $f(R)$ frame with fixed outer wedge $O^{f}_{W}[\sigma]$, then we maximize von Neumann entropy in the Einstein frame with fixed outer wedge $O^{E}_{W}[\sigma]$ at the same time. Finally, we find that the outer entropy of surface $\sigma$ are the same in $f(R)$ and the Einstein frame
\begin{equation}
    S^{\text{(outer)}}_{f}[\sigma]=S_{E}^{\text{(outer)}}[\sigma],
    \label{equivelence of outer entropy}
\end{equation}
if the outer wedge $O^{f}_{W}[\sigma]$ and $O^{E}_{W}[\sigma]$ are related by the transformation between the $f(R)$ frame and the Einstein frame.

In section \ref{section Coarse Gaining Entropy of Apparent Horizon in Einstein Frame}, we have already shown that in the Einstein frame, the outer entropy of the minimar surface $\mu$ is just the Bekenstein–Hawking entropy of the minimar surface
\begin{equation}
    S^{\text{(outer)}}_{E}[\mu]=\frac{\text{Area}_{E}[\mu]}{4G}.
\end{equation}
And from \eqref{equivelence of outer entropy}, we can write
\begin{equation}
    S^{\text{(outer)}}_{E}[\mu]=S_{f}^{\text{(outer)}}[\mu]=\frac{\text{Area}_{E}[\mu]}{4G}=\frac{\int_{\mu}f'(R)}{4G}.
\end{equation}
Recall that the minimar surface $\mu$ is defined by the vanishing of the outer null expansion, $\theta_{v_E}=0$, in the Einstein frame. Since null geodesics are conformally invariant, $k_E^a = (\partial_{v_E})^a$ remains a null geodesic in the $f(R)$ frame, but a different affine parameter $v$. The relationship between these parameters is given by \cite{Wald:1984rg}
\begin{equation}
    \frac{dv_{E}}{dv}=c\Omega^{2}.
\end{equation}
Therefore, we can show that the generalized expansion of $\mu$ in the $f(R)$ frame vanishes \cite{Kong:2024sqc}
\begin{equation}
    \theta_{v_{E}}=\frac{\partial}{\partial v^{E}}\text{log}\sqrt{\gamma^{E}}=\frac{1}{c\big(f'(R)\big)^{\frac{2}{D-2}}}\frac{\partial}{\partial v}\text{log}\big(\sqrt{\gamma}f'(R)\big)=\frac{1}{c\big(f'(R)\big)^{\frac{2}{D-2}}}\Theta_{v}=0,
\end{equation}
here $c$ is a positive constant, and this shows that $\Theta_{v}=0$. And the minimar condition of $\mu$ in the Einstein frame implies that $\mu$ satisfies the generalized minimar condition \footnote{It can be shown that $\nabla_{l_{E}}\theta^{E}_{k_{E}}|_{\mu}\propto \frac{f'(R)\partial_{u}\Theta_{v}-\Theta_{v}\partial_{u}f'(R)}{f'(R)^{2}}|_{\mu}=\frac{\partial_{u}\Theta_{v}}{f'(R)}|_{\mu}\leq0$. Since $f'(R)>0$, this implies $\nabla_{l}\Theta_{k}=\partial_{u}\Theta_{v}\leq 0$.} (see section \ref{Assumptions, Conventions, and Definitions}) in the $f(R)$ frame. Therefore, $\mu$ is just the generalized minimar surface in the $f(R)$ frame, and we finally show that the outer entropy of the generalized marginally trapped surface with generalized minimar condition in $f(R)$ gravity is just the Wald entropy of it
\begin{equation}
    S_{f}^{\text{(outer)}}[\mu]=\frac{\int_{\mu}f'(R)}{4G}.
    \label{section 4 outer entropy final result}
\end{equation}

\section{Deriving the Results in the \texorpdfstring{$f(R)$}{f(R)} Frame}
\label{Deriving the Results in f(R)}

In this section, we will derive the outer entropy directly in the $f(R)$ frame. To do this, we should first derive a focusing theorem for the generalized expansion $\Theta_{k}$ in $f(R)$ gravity. We will construct the  stationary null hypersurface $N_{-k}$ for the generalized expansion. Finally, we will construct the spacetime $(\mathcal{M}',g')$ that the von Neumann entropy that saturates the upper bound of the outer entropy.

\subsection{Nonlinear Raychaudhuri Equation For the Generalized Expansion}
\label{section 5.1}

Consider a null hypersurface $N_k$ generated by an affinely parametrized null geodesic vector field $k^a$, equipped with a "rigging" vector field $l^a$ satisfying $k_a l^a = -1$ (note that $l^a$ is not unique). The null and transverse extrinsic curvatures of $N_k$ are defined as
\begin{equation}
    \begin{split}
K^{(k)}_{ab}&=\gamma^{c}_{a}\gamma^{d}_{b}\nabla_{c}k_{d}\\K^{(l)}_{ab}&=\gamma^{c}_{a}\gamma^{d}_{b}\nabla_{c}l_{d},
    \end{split}
\end{equation}
here $\gamma_{ab}=g_{ab}+2l_{(a}k_{b)}$. And we define the expansion $\theta_{k}$ as
\begin{equation}
    \theta_{k}=\gamma^{ab}K^{(k)}_{ab}.
\end{equation}
Assuming $k^a$ is an affinely parametrized null geodesic, the Raychaudhuri equation for the expansion $\theta_k$ is given by
\begin{equation}
    k^{a}\nabla_{a}\theta_{k}=-\frac{1}{D-2}\theta_{k}^{2}-\varsigma^{ab}\varsigma_{ab}-R_{ab}k^{a}k^{b}.
\end{equation}
This equation is purely geometrical. 

For the $f(R)$ gravity, we define the generalized expansion $\Theta_{k}$
\begin{equation}
    \Theta_{k}=\gamma^{ab}K^{(k)}_{ab}+k^{a}\nabla_{a}\text{log}f'(R).
    \label{definition of generalized expansion}
\end{equation}
Assuming $k^a=(\partial_v)^a$ is affinely parametrized with affine parameter $v$, we can rewrite \eqref{definition of generalized expansion} as
\begin{equation}
    \Theta_{k}=\Theta_{v}=\partial_{v}\text{log}\big(\sqrt{\gamma}f'(R)\big).
\end{equation}
Using the equations of motion for $f(R)$ gravity \eqref{EOM of f(R)} and assuming $f'(R) > 0$, we find
\begin{equation}
    R_{ab}k^{a}k^{b}=\frac{8\pi GT_{ab}k^{a}k^{b}}{f'(R)}+\frac{1}{f'(R)}k^{a}k^{b}\nabla_{a}\nabla_{b}f'(R).
\end{equation}
We notice that 
\begin{equation}
    \frac{1}{f'(R)}k^{a}k^{b}\nabla_{a}\nabla_{b}f'(R)=k^{a}\nabla_{a}\big(k^{b}\nabla_{b}\text{log}f'(R)\big)+\big(k^{a}\nabla_{a}\text{log}f'(R)\big)^{2}.
\end{equation}
Substituting the $f(R)$ equations of motion into the Raychaudhuri equation yields
\begin{equation}
    k^{a}\nabla_{a}\Theta_{k}=-\frac{\theta_{k}^{2}}{D-2}-(k^{a}\nabla_{a}\text{log}f'(R))^{2}-\varsigma^{ab}\varsigma_{ab}-8\pi G\frac{T_{ab}k^{a}k^{b}}{f'(R)}.
    \label{generalized Raychaudhuri equation}
\end{equation}
Since we assume $f'(R)>0$, if the matter fields satisfy the NEC ($T_{ab}k^a k^b \ge 0$), we obtain a focusing theorem for the generalized expansion $\Theta_k$:
\begin{equation}
    k^{a}\nabla_{a}\Theta_{k}=\partial_{v}\Theta_{v}\leq0.
\end{equation}
We define \eqref{generalized Raychaudhuri equation} as the nonlinear Raychaudhuri equation.

We now use the nonlinear Raychaudhuri equation to calculate the upper bound of the outer entropy of the generalized marginally trapped surface. As discussed in section \ref{section Coarse Gaining Entropy of Apparent Horizon in Einstein Frame}, the NCC is satisfied in the Einstein frame. Therefore, under the AdS/CFT and some global assumptions, the von Neumann entropy $S_{\text{vN}}(\rho^{E\{\alpha\}}_{B})$ satisfies the maximin construction in the Einstein frame. Then we move into the $f(R)$ frame (two geometries are related by the Weyl transformation), this construction is equivalent to first finding the minimal Wald entropy surface $\min_f(B,\Sigma)$ homologous to $B$ on a given Cauchy slice $\Sigma$. We then choose $\Sigma$ to maximize the Wald entropy of $\min_f(B,\Sigma)$ over all possible Cauchy slices, the von Neumann entropy is the Wald entropy of $\min_f(B,\Sigma)$. 

Now consider the generalized marginally trapped surface $\mu$, let $N_{\pm k}(\mu)$ be the null congruence generated from $\mu$ by null geodesics $k^{a}$ and $-k^{a}$ in the $f(R)$ frame. Defining $\bar{\mu}$ as the representative of $\mu$ on Cauchy slice $\Sigma$, following the definition in section \ref{The Upper Bound of the Outer Entropy}. Assuming $\bar{\mu}$ lies on $N_k$, since $\Theta_v|_{\mu}=0$ and the nonlinear Raychaudhuri equation dictates $\partial_v\Theta_v \le 0$, it follows that
\begin{equation}
    \Theta_{v}|_{N_{k}}\leq0.
\end{equation}
In this case the Wald entropy of $\bar{\mu}$ is smaller than the Wald entropy of $\mu$, since 
\begin{equation}
    \partial_{v}\int_{\mathcal{T}}f'(R)dA=\int_{\mathcal{T}}\Theta_{v}f'(R)dA\leq0,
\end{equation}
here $\mathcal{T}$ is the cross section of $N_{k}$. If the representative $\bar{\mu}$ is on the $N_{-k}$, since $\partial_{v}\Theta_{v}\neq0$, it is easy to show that
\begin{equation}
     \Theta_{v}|_{N_{-k}}\geq0.
\end{equation}
Therefore, in this case the Wald entropy of $\bar{\mu}$ is smaller than the Wald entropy $\mu$, now we have already shown that
\begin{equation}
    \int_{\bar{\mu}}f'(R)\leq\int_{\mu}f'(R).
\end{equation}
Now consider a spacetime $(\mathcal{M}^{\{\alpha\}},g)$ with a fixed outer wedge $O^{f}_{W}[\mu]$. By the maximin construction \cite{Wall:2012uf}, there exist a Cauchy slice $\Sigma$ such that $X^{\{\alpha\}}$ (the surface that we calculate the von Neumann entropy in the $f(R)$ frame) has the minimal Wald entropy surface of $\Sigma$, therefore
\begin{equation}
    S_{\text{vN}}(\rho_{B}^{f})=\frac{\int_{X^{\{\alpha\}}}f'(R)}{4G}\leq\frac{\int_{\bar{\mu}}f'(R)}{4G}\leq\frac{\int_{\mu}f'(R)}{4G}.
\end{equation}
Since the outer entropy $S_f^{\text{(outer)}}$ is defined as the maximal von Neumann entropy, this implies
\begin{equation}
    S^{\text{(outer)}}_{f}[\mu]\leq\frac{\int_{\mu}f'(R)}{4G}.
\end{equation}

\subsection{Construction of Stationary Null Hypersurface For the Generalized Expansion}
\label{Construction of Stationary Null Hypersurface For the Generalized Expansion}

Similar to section \ref{section Coarse Gaining Entropy of Apparent Horizon in Einstein Frame}, constructing a spacetime with a fixed $O_W^f[\mu]$ that maximizes $S_{\rm vN}(\rho_B^f)$ requires gluing a stationary null hypersurface $N_{-k}$ for the generalized expansion, subject to specific initial data on $N_{-k}$. This will show that there exists a generalized extremal surface $X$ with zero generalized expansions of two null directions\footnote{In the Einstein frame, the surface $X$ where we evaluate the von Neumann entropy is the extremal surface. The extremal condition of $X$ in the Einstein frame will lead $X$ to a generalized extremal surface in the $f(R)$ frame.}, the Wald entropy of this surface is equal to the Wald entropy of $\mu$. The spacetime is then completed via a CPT reflection across $X$, analogous to the procedure in section \ref{section Coarse Gaining Entropy of Apparent Horizon in Einstein Frame} (see Figure \ref{CPT reflection}).

\begin{figure}[h]
    \centering

\tikzset{every picture/.style={line width=0.75pt}} 

\begin{tikzpicture}[x=0.75pt,y=0.75pt,yscale=-1,xscale=1]

\draw   (80.33,60) .. controls (81.56,61.71) and (82.73,63.33) .. (84.08,63.33) .. controls (85.44,63.33) and (86.61,61.71) .. (87.83,60) .. controls (89.06,58.29) and (90.23,56.67) .. (91.58,56.67) .. controls (92.94,56.67) and (94.11,58.29) .. (95.33,60) .. controls (96.56,61.71) and (97.73,63.33) .. (99.08,63.33) .. controls (100.44,63.33) and (101.61,61.71) .. (102.83,60) .. controls (104.06,58.29) and (105.23,56.67) .. (106.58,56.67) .. controls (107.94,56.67) and (109.11,58.29) .. (110.33,60) .. controls (111.56,61.71) and (112.73,63.33) .. (114.08,63.33) .. controls (115.44,63.33) and (116.61,61.71) .. (117.83,60) .. controls (119.06,58.29) and (120.23,56.67) .. (121.58,56.67) .. controls (122.94,56.67) and (124.11,58.29) .. (125.33,60) .. controls (126.56,61.71) and (127.73,63.33) .. (129.08,63.33) .. controls (130.44,63.33) and (131.61,61.71) .. (132.83,60) .. controls (134.06,58.29) and (135.23,56.67) .. (136.58,56.67) .. controls (137.94,56.67) and (139.11,58.29) .. (140.33,60) .. controls (141.56,61.71) and (142.73,63.33) .. (144.08,63.33) .. controls (145.44,63.33) and (146.61,61.71) .. (147.83,60) .. controls (149.06,58.29) and (150.23,56.67) .. (151.58,56.67) .. controls (152.94,56.67) and (154.11,58.29) .. (155.33,60) .. controls (156.56,61.71) and (157.73,63.33) .. (159.08,63.33) .. controls (160.44,63.33) and (161.61,61.71) .. (162.83,60) .. controls (164.06,58.29) and (165.23,56.67) .. (166.58,56.67) .. controls (167.94,56.67) and (169.11,58.29) .. (170.33,60) .. controls (171.56,61.71) and (172.73,63.33) .. (174.08,63.33) .. controls (175.44,63.33) and (176.61,61.71) .. (177.83,60) .. controls (179.06,58.29) and (180.23,56.67) .. (181.58,56.67) .. controls (182.94,56.67) and (184.11,58.29) .. (185.33,60) .. controls (186.56,61.71) and (187.73,63.33) .. (189.08,63.33) .. controls (190.44,63.33) and (191.61,61.71) .. (192.83,60) .. controls (194.06,58.29) and (195.23,56.67) .. (196.58,56.67) .. controls (197.94,56.67) and (199.11,58.29) .. (200.33,60) .. controls (201.56,61.71) and (202.73,63.33) .. (204.08,63.33) .. controls (205.44,63.33) and (206.61,61.71) .. (207.83,60) .. controls (209.06,58.29) and (210.23,56.67) .. (211.58,56.67) .. controls (212.94,56.67) and (214.11,58.29) .. (215.33,60) .. controls (216.56,61.71) and (217.73,63.33) .. (219.08,63.33) .. controls (220.44,63.33) and (221.61,61.71) .. (222.83,60) .. controls (224.06,58.29) and (225.23,56.67) .. (226.58,56.67) .. controls (227.94,56.67) and (229.11,58.29) .. (230.33,60) .. controls (231.56,61.71) and (232.73,63.33) .. (234.08,63.33) .. controls (235.44,63.33) and (236.61,61.71) .. (237.83,60) .. controls (239.06,58.29) and (240.23,56.67) .. (241.58,56.67) .. controls (242.94,56.67) and (244.11,58.29) .. (245.33,60) .. controls (246.56,61.71) and (247.73,63.33) .. (249.08,63.33) .. controls (250.44,63.33) and (251.61,61.71) .. (252.83,60) .. controls (254.06,58.29) and (255.23,56.67) .. (256.58,56.67) .. controls (257.94,56.67) and (259.11,58.29) .. (260.33,60) .. controls (261.56,61.71) and (262.73,63.33) .. (264.08,63.33) .. controls (265.44,63.33) and (266.61,61.71) .. (267.83,60) .. controls (269.06,58.29) and (270.23,56.67) .. (271.58,56.67) .. controls (272.94,56.67) and (274.11,58.29) .. (275.33,60) ;
\draw   (80.33,218.67) .. controls (81.56,220.37) and (82.73,222) .. (84.08,222) .. controls (85.44,222) and (86.61,220.37) .. (87.83,218.67) .. controls (89.06,216.96) and (90.23,215.33) .. (91.58,215.33) .. controls (92.94,215.33) and (94.11,216.96) .. (95.33,218.67) .. controls (96.56,220.37) and (97.73,222) .. (99.08,222) .. controls (100.44,222) and (101.61,220.37) .. (102.83,218.67) .. controls (104.06,216.96) and (105.23,215.33) .. (106.58,215.33) .. controls (107.94,215.33) and (109.11,216.96) .. (110.33,218.67) .. controls (111.56,220.37) and (112.73,222) .. (114.08,222) .. controls (115.44,222) and (116.61,220.37) .. (117.83,218.67) .. controls (119.06,216.96) and (120.23,215.33) .. (121.58,215.33) .. controls (122.94,215.33) and (124.11,216.96) .. (125.33,218.67) .. controls (126.56,220.37) and (127.73,222) .. (129.08,222) .. controls (130.44,222) and (131.61,220.37) .. (132.83,218.67) .. controls (134.06,216.96) and (135.23,215.33) .. (136.58,215.33) .. controls (137.94,215.33) and (139.11,216.96) .. (140.33,218.67) .. controls (141.56,220.37) and (142.73,222) .. (144.08,222) .. controls (145.44,222) and (146.61,220.37) .. (147.83,218.67) .. controls (149.06,216.96) and (150.23,215.33) .. (151.58,215.33) .. controls (152.94,215.33) and (154.11,216.96) .. (155.33,218.67) .. controls (156.56,220.37) and (157.73,222) .. (159.08,222) .. controls (160.44,222) and (161.61,220.37) .. (162.83,218.67) .. controls (164.06,216.96) and (165.23,215.33) .. (166.58,215.33) .. controls (167.94,215.33) and (169.11,216.96) .. (170.33,218.67) .. controls (171.56,220.37) and (172.73,222) .. (174.08,222) .. controls (175.44,222) and (176.61,220.37) .. (177.83,218.67) .. controls (179.06,216.96) and (180.23,215.33) .. (181.58,215.33) .. controls (182.94,215.33) and (184.11,216.96) .. (185.33,218.67) .. controls (186.56,220.37) and (187.73,222) .. (189.08,222) .. controls (190.44,222) and (191.61,220.37) .. (192.83,218.67) .. controls (194.06,216.96) and (195.23,215.33) .. (196.58,215.33) .. controls (197.94,215.33) and (199.11,216.96) .. (200.33,218.67) .. controls (201.56,220.37) and (202.73,222) .. (204.08,222) .. controls (205.44,222) and (206.61,220.37) .. (207.83,218.67) .. controls (209.06,216.96) and (210.23,215.33) .. (211.58,215.33) .. controls (212.94,215.33) and (214.11,216.96) .. (215.33,218.67) .. controls (216.56,220.37) and (217.73,222) .. (219.08,222) .. controls (220.44,222) and (221.61,220.37) .. (222.83,218.67) .. controls (224.06,216.96) and (225.23,215.33) .. (226.58,215.33) .. controls (227.94,215.33) and (229.11,216.96) .. (230.33,218.67) .. controls (231.56,220.37) and (232.73,222) .. (234.08,222) .. controls (235.44,222) and (236.61,220.37) .. (237.83,218.67) .. controls (239.06,216.96) and (240.23,215.33) .. (241.58,215.33) .. controls (242.94,215.33) and (244.11,216.96) .. (245.33,218.67) .. controls (246.56,220.37) and (247.73,222) .. (249.08,222) .. controls (250.44,222) and (251.61,220.37) .. (252.83,218.67) .. controls (254.06,216.96) and (255.23,215.33) .. (256.58,215.33) .. controls (257.94,215.33) and (259.11,216.96) .. (260.33,218.67) .. controls (261.56,220.37) and (262.73,222) .. (264.08,222) .. controls (265.44,222) and (266.61,220.37) .. (267.83,218.67) .. controls (269.06,216.96) and (270.23,215.33) .. (271.58,215.33) .. controls (272.94,215.33) and (274.11,216.96) .. (275.33,218.67) ;
\draw    (275.33,59.33) -- (275.33,219.33) ;
\draw    (80.67,60.67) -- (80.67,220) ;
\draw [color={rgb, 255:red, 208; green, 2; blue, 27 }  ,draw opacity=1 ]   (209,113.67) -- (274.67,180.67) ;
\draw [color={rgb, 255:red, 208; green, 2; blue, 27 }  ,draw opacity=1 ]   (80.33,99) -- (146,166) ;
\draw  [dash pattern={on 0.84pt off 2.51pt}]  (80.67,60.67) -- (275.33,219.33) ;
\draw [color={rgb, 255:red, 18; green, 16; blue, 224 }  ,draw opacity=1 ]   (209,113.67) -- (230.36,96.26) ;
\draw [shift={(231.91,95)}, rotate = 140.83] [color={rgb, 255:red, 18; green, 16; blue, 224 }  ,draw opacity=1 ][line width=0.75]    (10.93,-3.29) .. controls (6.95,-1.4) and (3.31,-0.3) .. (0,0) .. controls (3.31,0.3) and (6.95,1.4) .. (10.93,3.29)   ;
\draw [color={rgb, 255:red, 18; green, 16; blue, 224 }  ,draw opacity=1 ]   (209,113.67) -- (188.02,92.81) ;
\draw [shift={(186.6,91.4)}, rotate = 44.83] [color={rgb, 255:red, 18; green, 16; blue, 224 }  ,draw opacity=1 ][line width=0.75]    (10.93,-3.29) .. controls (6.95,-1.4) and (3.31,-0.3) .. (0,0) .. controls (3.31,0.3) and (6.95,1.4) .. (10.93,3.29)   ;
\draw [color={rgb, 255:red, 208; green, 2; blue, 27 }  ,draw opacity=1 ]   (146,166) -- (209,113.67) ;
\draw    (231.91,95) -- (275.33,59.33) ;
\draw    (146,166) -- (80.67,220) ;

\draw (178,139.67) node  [font=\Large]  {$\cdot $};
\draw (209,113.67) node  [font=\Large]  {$\cdot $};
\draw (146,166) node  [font=\Large]  {$\cdot $};
\draw (211.56,128.2) node  [font=\scriptsize]  {$\mu $};
\draw (149.56,179.53) node  [font=\scriptsize]  {$\tilde{\mu }$};
\draw (180.89,155.53) node  [font=\scriptsize]  {$X$};
\draw (242.89,114.2) node  [font=\scriptsize]  {$O_{W}^{f}[ \mu ]$};
\draw (225.56,89.8) node  [font=\scriptsize,color={rgb, 255:red, 19; green, 23; blue, 254 }  ,opacity=1 ]  {$k$};
\draw (199.16,90.2) node  [font=\scriptsize,color={rgb, 255:red, 19; green, 23; blue, 254 }  ,opacity=1 ]  {$l$};
\draw (140.78,94.3) node  [font=\scriptsize,rotate=-39.48] [align=left] {CPT reflection};
\draw (253.24,179.14) node  [font=\scriptsize]  {$N_{-l}[ \mu ]$};
\draw (200.74,145.14) node  [font=\scriptsize]  {$N_{-k}[ \mu ]$};
\draw (105.24,151.64) node  [font=\scriptsize]  {$N'_{-l}\left[\tilde{\mu }\right]$};
\draw (64.74,141.14) node  [font=\scriptsize]  {$\tilde{B}$};
\draw (291.24,138.14) node  [font=\scriptsize]  {$B$};

\end{tikzpicture}
\caption{This is the full maximizing spacetime in the $f(R)$ frame. The characteristic Cauchy slice that we construct to obtain this geometry consists of $\Sigma=N_{-l}[\mu]\cup N_{-k}[\mu]\cup N'_{-l}[\tilde{\mu}]$.}
\label{CPT reflection}
\end{figure}

We now first construct the stationary null hypersurface $N_{-k}$ for generalized expansion. Adopting the same gauge as in section \ref{section Coarse Gaining Entropy of Apparent Horizon in Einstein Frame}, we use null coordinates $u, v$ and spatial coordinates $x^i$, fixing $\mu$ at $u=v=0$ and placing $N_{-k}$ at $u=0$. $l^{a}=(\partial_{u})^{a}$ and $k^{a}=(\partial_{v})^{a}$ are the generating null vectors of $N_{-l}$ and $N_{-k}$. We use the GNC gauge, that is,
\begin{equation}
\begin{split}
g_{uv}&=-1\\g_{uu}&=g_{ui}=0\\g_{vv}|_{u=0}=g_{vv,u}&|_{u=0}=g_{vi}|_{u=0}=0.
\end{split}
\end{equation}
We can write the line element under these gauge conditions (GNC) \cite{Wall:2015raa, Wall:2024lbd, Hollands:2022fkn, Bhattacharyya:2021jhr, Visser:2024pwz, Hollands:2024vbe}
\begin{equation}
    ds^{2}=-2dudv-u^{2}\alpha dv^{2}-2u\omega_{i}dvdx^{i}+\gamma_{ij}dx^{i}dx^{j}.
\end{equation}
In this gauge, the twist and the null extrinsic curvatures are
\begin{equation}
    2\chi_{i}=g_{iv,u}=\omega_{i}+u\partial_{u}\omega_{i}
    \label{twist in f(R)}
\end{equation}
\begin{equation}
    2K_{ij}^{(l)}=g_{ij,u}
\end{equation}
\begin{equation}
    2K_{ij}^{(k)}=g_{ij,v}|_{u=0}.
\end{equation}

In this gauge, the constraint equations on the stationary null hypersurface $N_{-k}$ for twist and generalized expansions reduce to \cite{Engelhardt:2018kcs, Hayward:2006ss, Hayward:2001ck, Hayward:2004fz, Gourgoulhon:2005ng, Cao:2010vj}
\begin{equation}
    \Theta_{u,v}=-\frac{1}{2}\mathcal{R}+\nabla\cdot\chi-\theta_{u}\theta_{v}+\partial_{v}\partial_{u}\text{log}f'(R)+R_{uv}+\chi^{2}
\end{equation}
\begin{equation}
    \Theta_{v,v}=-\frac{\theta_{k}^{2}}{D-2}-(\partial_{v}\text{log}f'(R))^{2}-\varsigma_{v}^{2}-8\pi G\frac{T_{vv}}{f'(R)}
    \label{generalized Raychaudhuri equation on Nk}
\end{equation}
\begin{equation}
    \chi_{i,v}=-\theta_{v}+\Big(\frac{D-3}{D-2}\Big)\nabla_{i}\theta_{v}-(\nabla\cdot\varsigma_{v})_{i}+R_{iv}.
\end{equation}
We now specify the initial data on $N_{-k}$. Based on our analysis in the Einstein frame, $\partial_v \log f'(R)$ represents free data that can be set to zero. Since in the Einstein frame, $\partial_{v_E}\phi$ is a free data on the null hypersurface $N_{-k_{E}}$
\begin{equation}
    \partial_{v_{E}}\phi=\frac{1}{\sqrt{16\pi G}}\sqrt{\frac{2(D-1)}{D-2}}\ \partial_{v_{E}}\text{log}f'(R)\propto\partial_{v}\text{log}f'(R).
\end{equation}
Therefore, in the $f(R)$ frame $\partial_{v}\text{log}f'(R)$ should be free data that we set it to zero. Since we assume $f'(R)>0$, this is equivalent to $\partial_{v}f'(R)=0$. We will require
\begin{align}
    \partial_{v}f'(R)[N_{-k}]&=0\\\varsigma_{v}[N_{-k}]&=0\label{varsigma on Nk}\\ T_{vv}[N_{-k}]&=0\\T_{iv}[N_{-k}]&=0\\T_{uv}[N_{-k}]&=\text{const}\label{initial free data Tuv}\\T_{ij}[N_{-k}]&=\text{const}\label{initial free data Tij}.
\end{align}
Since $\partial_{v}\text{log}f'(R)=0$, it is easy to show that $\Theta_{v}[N_{-k}]=\theta_{v}[N_{-k}]$. Substituting the first three equalities into \eqref{generalized Raychaudhuri equation on Nk} reveals that $\Theta_v = \theta_v = 0$ on $N_{-k}$, implying that $R$ is also constant along $N_{-k}$. Now we consider the quantity $R_{iv}$, from the EOM of $f(R)$ gravity, we can show that
\begin{equation}
    R_{iv}=\frac{1}{2}g_{iv}\frac{f(R)}{f'(R)}+\frac{1}{f'(R)}(\nabla_{i}\nabla_{v}-g_{iv}\nabla^{2})f'(R)+8\pi G\frac{T_{iv}}{f'(R)}.
\end{equation}
Since $g_{iv}|_{N_{-k}}=T_{iv}[N_{-k}]=0$, we only need to evaluate $\frac{1}{f'(R)}\nabla_{i}\nabla_{v}f'(R)$. One can show that \cite{Visser:2024pwz}
\begin{equation}
    \nabla_{i}\nabla_{v}f'(R)|_{u=0}=\partial_{i}\partial_{v}f'(R)|_{u=0}-K^{j}_{i}\partial_{j}f'(R)|_{u=0}=0,
\end{equation}
we already use that on $N_{-k}$, $\theta_{v}=\varsigma_{v}=0$, this will imply $K^{j}_{i}[N_{-k}]=0$. Therefore, we have shown that on the stationary null hypersurface $N_{-k}$
\begin{equation}
    R_{iv}[N_{-k}]=0.
\end{equation}
Combined with \eqref{varsigma on Nk}, this shows that $\chi_{i,v} = 0$, which restricts the twist to be constant along $N_{-k}$. From \eqref{twist in f(R)}, we find that 
\begin{equation}
    \chi_{i,v}[N_{-k}]=\frac{1}{2}\partial_{v}\omega_{i}[N_{-k}]=0,
\end{equation}
therefore, on the stationary null hypersurface $N_{-k}$, $\omega_{i}$ is constant, that is, $\partial_{v}\omega_{i}=0$.

Next, consider $\Theta_{u,v}$. Since $\theta_v=\varsigma_v=0$ on the stationary null hypersurface $N_{-k}$, it follows that $\partial_v \gamma_{ij} = 0$. Therefore, $\mathcal{R}$ will be a constant on $N_{-k}$. Since we have already shown that $\chi_i$ is constant and $\theta_v = 0$ on $N_{-k}$, the combination $\partial_v\partial_u \log f'(R) + R_{uv}$ is the only term that could potentially be non-constant on $N_{-k}$. Now we use the EOM of $f(R)$ gravity \eqref{EOM of f(R)} to show that this term is also a constant on $N_{-k}$. We first consider $R_{uv}$
\begin{equation}
    R_{uv}=\frac{1}{2}g_{uv}\frac{f(R)}{f'(R)}+\frac{1}{f'(R)}(\nabla_{u}\nabla_{v}-g_{uv}\nabla^{2})f'(R)+8\pi G\frac{T_{uv}}{f'(R)}.
\end{equation}
Since $f'(R)$ and $g_{uv}$ are constant on $N_{-k}$, we must determine whether $f(R)$ is also constant. We know that $\partial_{v}f(R)=\partial_{v}R\cdot f'(R)$, since $\partial_{v}f'(R)=\partial_{v}R\cdot f''(R)=0$ and for simplicity we assume $f''(R)\neq0$ \cite{Kong:2024sqc} (in Appendix \ref{Cases of f''(R)=0} we will discuss the cases of $f''(R)=0$), this will show that $\partial_{v}R=0$. Therefore, $f(R)$ is constant on $N_{-k}$. And we know that $T_{uv}$ is a constant on $N_{-k}$, then $8\pi G(T_{uv}/f'(R))$ is also a constant. Finally, we consider the remaining terms
\begin{equation}
    \frac{1}{f'(R)}\partial_{v}\partial_{u}f'(R)+\frac{1}{f'(R)}(\nabla_{u}\nabla_{v}-g_{uv}\nabla^{2})f'(R).
\end{equation}
It is easy to show that \cite{Visser:2024pwz}
\begin{equation}
    \nabla_{u}\nabla_{v}f'(R)|_{u=0}=\partial_{u}\partial_{v}f'(R)+\frac{1}{2}\omega^{i}\partial_{i}f'(R).
\end{equation}
And we can show that 
\begin{equation}
\begin{split}
\nabla^{2}f'(R)|_{u=0}=-2\partial_{u}\partial_{v}f'&(R)+\Delta_{\gamma}f'(R)-\partial_{v}\text{log}\sqrt{\gamma}\partial_{u}f'(R)\\-\partial_{u}\text{log}\sqrt{\gamma}\partial_{v}f'(R)-\omega^{i}\partial_{i}f'(R)&=-2\partial_{u}\partial_{v}f'(R)+\Delta_{\gamma}f'(R)-\omega^{i}\partial_{i}f'(R),
\end{split}
\end{equation}
here $\Delta_{\gamma}$ is the laplacian of $x^{i}$. Then the remaining terms are reduced to
\begin{equation}
\begin{split}
     \frac{1}{f'(R)}&\partial_{v}\partial_{u}f'(R)+\frac{1}{f'(R)}(\nabla_{u}\nabla_{v}-g_{uv}\nabla^{2})f'(R)\\&=\frac{1}{f'(R)}\Big(\Delta_{\gamma}f'(R)-\frac{1}{2}\omega^{i}\partial_{i}f'(R)\Big).
\end{split}
\end{equation}
Since $\partial_{v}\gamma_{ij}=0$ and $\chi_{i}$ is constant on $N_{-k}$, it can be shown that $\Delta_{\gamma}f'(R)$ and $\omega^{i}\partial_{i}f'(R)$ are constant on $N_{-k}$. Finally, we show that $R_{iv}$ is a constant on $N_{-k}$, this will lead to $\Theta_{u,v}$ is a constant on $N_{-k}$.

We now need to show that our construction satisfies the junction conditions. Since the junction conditions have beautiful forms in the Einstein frame, we now first move back to the Einstein frame. Given that we have set $\partial_v f'(R) = 0$, one can easily show that $k^a|_{N_{-k}} = (\partial_v)^a|_{N_{-k}}$ remains an affinely parametrized null geodesic \cite{Wald:1984rg}
\begin{equation}
    k^{a}\nabla_{a}^{E}k^{b}|_{N_{-k}}=(2k^{a}\nabla^{E}_{a}\text{log}\Omega)\cdot k^{a}=0,
\end{equation}
here $\Omega=\big(f'(R)\big)^{\frac{1}{D-2}}$ and we already use $k^{a}\nabla_{a}\Omega=\partial_{v}\Omega=0$. Therefore, it is reasonable to define $k^{a}_{E}|_{N_{-k}}=(\partial_{v_{E}})^{a}|_{N_{-k}}=(\partial_{v})^{a}|_{N_{-k}}$. Now let us define the ''rigging" vector field $l^{a}_{E}$. Weyl transformation preserves the null geodesic and the relation between affine parameter before and after Weyl transformation is \cite{Wald:1984rg}
\begin{equation}
    \frac{du_{E}}{du}=c\Omega^{2},
\end{equation}
here $c$ is a positive constant and now we choose $c=1$. Therefore, $l_{E}^{a}=(\partial_{u_{E}})^{a}=(1/\Omega^{2})(\partial_{u})^{a}$, and we define $N_{-k}$ is located at $u_{E}=0$. This will guarantee that on $N_{-k}$
\begin{equation}
    g^{E}_{ab}k_{E}^{a}l_{E}^{b}=\frac{1}{\Omega^{2}}\Omega^{2}g_{ab}k^{a}l^{b}=-1,
\end{equation}
with $(\partial_{u_{E}})^{a}$ is affine null geodesic everywhere under the metric $g^{E}_{ab}=\Omega^{2}g_{ab}$. Since Weyl transformation preserves the orthogonal relationship between $l$, $k$ and $x$. On $N_{-k}$, we define $(x^{E}_{i})^{a}=(\partial_{x^{i}_{E}})^{a}=(\partial_{x^{i}})^{a}$. We now extend $v_{E}$ and $x^{i}_{E}$ by keeping the parameter $v_{E}$ and $x^{i}_{E}$ fixed along the null geodesics generated by $l_{E}^{a}$, and demanding that $k_{E}^{a}=(\partial_{v_{E}})^{a}$ and $(x^{E}_{i})^{a}=(\partial_{x^{i}_{E}})^{a}$ everywhere \cite{Visser:2024pwz}
\begin{equation}
   [k_{E},l_{E}]^{a}=[x_{i}^{E},l_{E}]^{a}=0,
\end{equation}
and the line element of the Einstein frame 
\begin{equation}
    ds^{2}_{E}=-2du_{E}dv_{E}-u_{E}^{2}\alpha_{E}dv_{E}^{2}-2u_{E}\omega^{E}_{i}dv_{E}dx^{i}_{E}+\gamma^{E}_{ij}dx^{i}_{E}dx^{i}_{E}.
\end{equation}
This implies that we reconstruct a GNC coordinate near $N_{-k}$ in the Einstein frame. In the Einstein frame, the junction conditions are \eqref{junction conditions in the Einstein frame}. Since $\theta^{E}_{v_{E}}[\mu]=\Theta_{v}[\mu]=\theta^{E}_{v_{E}}[N_{-k}]=\Theta_{v}[N_{-k}]=0$, the first junction condition is satisfied automatically. Now for the second junction condition, $\theta^{E}_{u_{E},v_{E}}[N_{-k}]=(1/\Omega^{2})\Theta_{u,v}[N_{-k}]$ is obviously a constant along $N_{-k}$. Since $\theta^{E}_{u_{E},v_{E}}[N_{-k}]$ is defined up to functions of the transverse $x^{i}_{E}|_{N_{-k}}=x^{i}|_{N_{-k}}$ directions. Even after fixing $\theta_{u_E, v_E}^E[N_{-k}]$, we retain the freedom to choose $\theta_{u_E}^E$ at $v=v_E=0$ on $N_{-k}$. For $\chi_{i}^{E}$, we have
\begin{equation}
\chi_{i}^{E}[N_{-k}]=\partial_{u_{E}}g^{E}_{v_{E}i}|_{N_{-k}}=\omega^{E}_{i}[N_{-k}]=\frac{1}{\Omega^{2}}\partial_{u}(u\Omega^{2}\omega_{i})|_{N_{-k}}=\omega_{i}[N_{-k}], 
\end{equation}
remains defined up to functions of the transverse $x^{i}_{E}|_{N_{-k}}=x^{i}|_{N_{-k}}$ directions, we are always free to choose all of these quantities to be continuous across the junction at $\mu$
\begin{align}
     [\theta^{E}_{v_{E}}]&=0\\ [\theta^{E}_{u_{E}}]&=0\\ [\chi^{E}_{i}]&=0.
\end{align}
Similarly, since $\phi$ and the matter field $\Phi$ are defined up to transverse functions, we are always free to enforce these quantities to be continuous across the junction at $\mu$, i.e., $[\phi]=[\Phi]=0$. These choices satisfy the junction conditions for the metric \eqref{junction conditions in the Einstein frame} and for the matters. This implies that we will get a continuous spacetime and well defined stress tensor across $\mu$ in the Einstein frame, and therefore in the $f(R)$ frame.

Therefore, we glue a stationary null hypersurface $N_{-k}$ for the generalized expansion $\Theta_{k}$ to $\mu$. Our next step is to find the generalized extremal surface $X$ with zero generalized expansions of two null directions.

\subsection{Construction of the Generalized Extremal Surface and the Outer Entropy in the \texorpdfstring{$f(R)$}{f(R)} Frame}

We are going to find the generalized extremal surface $X$ on $N_{-k}$. Since the generalized expansion $\Theta_k[N_{-k}] = 0$ on the stationary null hypersurface $N_{-k}$, we only need to find a surface or a cross-section of $N_{-k}$ on which $\Theta_{l}=0$. Similarly to section \ref{Construct the Geometry that Saturates the Upper Bound}, on a constant $v$ slice $\Theta_{u,v}=\text{const}$ and from the condition of the generalized minimal surface $\Theta_{u,v}<0$. This will ensure that there exists a surface on $N_{-k}$ with $\Theta_{l}=0$, but this surface does not need to be a constant $v$ slice. Consider a slice $\beta$ with varying $v$, defined by $v = h(x^i)$. By definition, $k^a = (\partial_v)^a$ is normal to $\beta$. We define the second null normal to $\beta$, denote by $w^{a}$, it can be shown that \cite{Engelhardt:2018kcs}
\begin{equation}
    w^{a}=l^{a}+\sum_{i}x_{i}^{a}\nabla_{i}h+\frac{1}{2}k^{a}\Delta_{\gamma} h,
\end{equation}
here $\nabla_{i}=x_{i}^{a}\nabla_{a}$. The generalized expansion of $w^{a}$ is defined as $\Theta_{w}=\theta_{w}+w^{a}\nabla_{a}\text{log}f'(R)$, it can be shown that \cite{Engelhardt:2018kcs}
\begin{equation}
    \Theta_{w}[\beta]=\Theta_{u}[\beta]+\Delta_{\gamma} h(x^{i})+(2\chi+\nabla\text{log}f'(R))\cdot\nabla h(x^{i}).
\end{equation}
Since we already show that $\Theta_{u,v}$ is independent of $v$ on constant $v$ slice, we can show that \cite{Andersson:2005gq, Andersson:2007fh}
\begin{equation}
\begin{split}
    \Theta_{w}[\beta]&=\Theta_{u}[\mu]+\Theta_{u,v}[\mu]h(x^{i})+(2\chi+\nabla\text{log}f'(R))\cdot\nabla h(x^{i})+\Delta_{\gamma} h(x^{i})\\&=L^{\mu}[h]+\Theta_{u}[\mu].
\end{split}
\end{equation}
Since $\chi$ and $\gamma$ are independent of $v$, the linear operator $L^{\mu}$ depends only on quantities evaluated at $\mu$, this linear operator is known as the stability operator \cite{Andersson:2005gq, Andersson:2007fh}, and now we generalized it to $f(R)$ gravity.

To locate the generalized extremal surface $X$ where $\Theta_w[X] = 0$, we must solve the second-order PDE:
\begin{equation}
    L^{\mu}[h]=-\Theta_{u}[\mu].
\end{equation}
From the results of 2-order PDE on a closed manifold (co-dimension 2 surface $\mu$), if $\Theta_{u,v}[\mu]<0$ and $\Theta_{u}[\mu]>0$, this 2-order PDE has a nontrivial  solution and $h(x^{i})<0$ \cite{Andersson:2005gq, Andersson:2007fh, Engelhardt:2018kcs}. Finally, we conclude that there exists a generalized extremal surface $X$ on $N_{-k}$ with two null generalized expansions zero. Since $N_{-k}$ is stationary for the generalized expansion $\Theta_{k}$,
\begin{equation}
    \int_{X}f'(R)=\int_{\mu}f'(R).
\end{equation}

Similarly to section \ref{Construct the Geometry that Saturates the Upper Bound}, we complete our construction by a CPT-reflection across $X$ that takes $v\to-v$, $u\to-u$, $x^i\to x^{i}$. Under the CPT transformation, $\Phi$, $f'(R)$ and $\chi_{i}$ are even, whereas $\Theta_{u}$, $\Theta_{v}$ and $\partial_{v}f'(R)$ are odd. All quantities that are odd under CPT vanish on $X$ by construction. Therefore, the CPT-conjugate data satisfies the requisite junction conditions. The results is a second boundary $\tilde{B}$, $B$ and $\tilde{B}$ are connected by a Cauchy slice with three null segments $\Sigma=N_{-l}[\mu]\cup N_{-k}[\mu]\cup N'_{-l}[\tilde{\mu}]$, we are using prime to represent CPT-conjugated submanifolds, see figure \ref{CPT reflection}. We have now specified all data necessary to uniquely evolve characteristic initial data via the EOM of $f(R)$ gravity. The resulting spacetime $(\mathcal{M}',g)$ has a generalized minimal surface $\mu$ with a fixed outer wedge $O_{W}^{f}[\mu]$. This spacetime contains a generalized extremal surface $X$ on the boundary of the inner wedge $I_{W}^{f}[\mu]$, which is homologous to $\mu$ and therefore to the boundary $B$. 

Finally, we need to show that this generalized extremal surface $X$ has the least Wald entropy. Given the duality established in \eqref{von Neumann entropy correspondence}, $X$ must have a minimal area in the Einstein frame if $X$ is the extremal surface. Therefore, in the $f(R)$ frame, $X$ should have the least Wald entropy if there exist other generalized extremal surfaces. If the constructed spacetime $(\mathcal{M}', g')$ contains no other generalized extremal surfaces, the proof is complete. If there exists other generalized extremal surfaces we denote it by $X'$, define $\overline{X'}[\Sigma]$ as the representative of $X'$ on Cauchy slice $\Sigma=N_{-l}[\mu]\cup N_{-k}[\mu]\cup N'_{-l}[\tilde{\mu}]$. There are in total three cases; If $\overline{X'}[\Sigma]$ lies entirely on $N_{-k}$, then, because $N_{-k}$ is stationary with respect to the generalized expansion, we have
\begin{equation}
     \int_{\overline{X'}[\Sigma]}f'(R)=\int_{X}f'(R).
\end{equation}

If $\overline{X'}[\Sigma]$ lies on $N_{-l}[\mu]$ or $N'_{-l}[\tilde{\mu}]$, from the focusing theorem \eqref{generalized Raychaudhuri equation} for $f(R)$ gravity
\begin{equation}
    \int_{\overline{X'}[\Sigma]}f'(R)\geq\int_{X}f'(R).
\end{equation}
And for the same reason, in both cases 
\begin{equation}
    \int_{X'}f'(R)\geq\int_{\overline{X'}[\Sigma]}f'(R)\geq\int_{X}f'(R).
\end{equation}

If $\overline{X'}[\Sigma]$ lies on $N_{-l}[\mu]$ and $N_{-k}$, define $s_{1}=\overline{X'}[\Sigma]\cap N_{-l}[\mu]$, $s_{2}=\overline{X'}[\Sigma]\cap N_{-k}[\mu]$, $\mu_{1}=\mu\cap O_{W}^{f}[X']$ and $\mu_{2}$ as the complement in $\mu$. It is easy to show that $\mu_{1}\cup s_{1}$ and $\mu_{2}\cup s_{2}$ homologous to $\mu$, then by the focusing theorem
\begin{align}
\int_{s_{1}}f'(R)+\int_{\mu_{1}}f'(R)\geq\int_{\mu}f'(R)\label{a}\\\int_{s_{2}}f'(R)+\int_{\mu_{2}}f'(R)=\int_{\mu}f'(R)\label{b}.
\end{align}
Then we can combine \eqref{a} and \eqref{b}, we find
\begin{equation}
    \int_{X'}f'(R)\geq\int_{\overline{X'}[\Sigma]=s_{1}\cup s_{2}}f'(R)\geq\int_{\mu}f'(R)=\int_{X}f'(R).
\end{equation}

Therefore, the preceding discussion establishes that $X$ minimizes the Wald entropy among all generalized extremal surfaces in $(\mathcal{M}', g)$. Consequently, for the spacetime $(\mathcal{M}', g)$,
\begin{equation}
    S_{\text{vN}}(\rho')=\frac{\int_{X}f'(R)}{4G}
\end{equation}
which saturates the upper bound of the outer entropy. This shows that
\begin{equation}
    S^{\text{(outer)}}_{f}[\mu]=\frac{\int_{\mu}f'(R)}{4G}.
\end{equation}

\section{The Boundary Dual of the Outer Entropy in \texorpdfstring{$f(R)$}{f(R)} Gravity}
\label{The Boundary Dual of the Outer Entropy}

Our constructions so far have focused on the bulk side constructions. To achieve a fully holographic generalization of the coarse-grained entropy for a black hole in $f(R)$ gravity, we must define a dual quantity on the boundary, namely the simple entropy \cite{Engelhardt:2018kcs, Engelhardt:2017aux}. In this section, we demonstrate that in $f(R)$ gravity, the simple entropy serves as the exact boundary dual to the outer entropy of the generalized marginally trapped surface. Furthermore, we discuss the second law as applied to both the simple and outer entropy.

The simple entropy is defined via the coarse-graining of $S_{\rm vN}$, holding fixed the expectation values of ``simple" boundary operators in the presence of all possible ``simple" sources. Sources are defined as ``simple" if the bulk fields they generate propagate causally into the bulk spacetime\footnote{It is worth noting that in $f(R)$ gravity there exists a dynamical operator $f'(R)$ in the bulk, which satisfy the Klein-Gordan equation $\Box f'(R)+V(f'(R))=\big(8\pi G/(D-1)\big)T$. Therefore, $f'(R)$ operator propagate causally in the bulk spacetime, that is, a simple operator.}; correspondingly, operators are ``simple" if their associated sources are simple \cite{Engelhardt:2018kcs}. The coarse-graining procedure consists of three steps. First, we select a boundary initial time $t_i$ and a late-time cutoff $t_f$. Second, we fix all one-point functions of the local operators for times $t > t_i$, in the presence of all possible simple sources after $t_i$. Finally, find the state $\rho$ that maximizes the von Neumann entropy, more precisely \cite{Engelhardt:2018kcs,Engelhardt:2017aux}
\begin{equation}
    S^{\text{(simple)}}[t_{i}]=\max_{\rho}\Big[S_{\text{vN}}(\rho)\Big|\text{ fixed }\langle E\mathcal{O}E^{\dagger}\rangle_{\rho}\Big],
\end{equation}
where $E$ represents the presence of all possible simple sources.

\subsection{Evaluating the Simple Entropy}

To establish that the simple entropy is the correct boundary dual to the outer entropy, we first demonstrate that the outer entropy acts as the lower bound for the simple entropy. The first step is relating the simple entropy $S^{\text{(simple)}}[t_{i}]$ to the outer entropy of a bulk surface $\mu$. Consider the slice $t_{i}$ and shoot a future direct null hypersurface $N_{l}[t_{i}]=\partial I^{+}[t_{i}]$. For an $f(R)$ black hole, there must exist a surface $\mu$ on $N_l[t_i]$ with vanishing $\Theta_k$. For the outer most $\mu$, the condition $\nabla_{l}\Theta_{k}<0$\footnote{This is the generalized minimar condition, see section \ref{Assumptions, Conventions, and Definitions}.} should be generically satisfied. By the focusing theorem in $f(R)$ gravity \eqref{generalized Raychaudhuri equation}, $\mu$ minimizes the Wald entropy on $N_l[t_i]$ relative to any surface outside it. Therefore, $\mu$ is the generalized minimar surface.

To compare the simple entropy and the outer entropy, it is natural to compare the information that is fixed under the change of state $\rho$. More precisely, compare the bulk region that can be reconstructed by the one point data after $t_{i}$ with the outer wedge of $\mu$. For the fixed simple source $J$, following the HKLL prescription\footnote{It is also possible to include interactions, at least perturbatively in 1/N \cite{Kabat:2011rz, Heemskerk:2012mn}.} \cite{Hamilton:2005ju, Hamilton:2006az, Hamilton:2006fh} and EOM in the bulk, we can use all one-point data after $t_{i}$, namely $\langle E\mathcal{O}E^{\dagger}\rangle$, to reconstruct $R_{J}[t_{i}]=D[I^{+}[t_{i}]\cap I^{-}[\partial\mathcal{M}]]$. The information outside the $R_{J}[t_{i}]$ cannot be reconstructed from one-point data after $t_{i}$. More precisely, since we only consider the simple operators which propagate causally in bulk, the modification outside the $R_{J}[t_{i}]$ cannot affect boundary operators after $t_{i}$.

We now apply arbitrary simple sources after $t_{i}$, since sources are simple, the change of bulk geometry should be constrained in $I^{+}[t_{i}]$ which does not affect $N_{l}[t_{i}]$. Therefore, when we need to compare $R_{J}$ and $R_{J'}$, we can compare the part of $N_{l}[t_{i}]$ contained in $R_{J}$ and $R_{J'}$. When the simple source $J$ is turned on, we either introduce or remove infalling matter. This will make the event horizon $I^{-}[\partial\mathcal{M}]$ move outward or inward along $N_{l}[t_{i}]$. In Einstein gravity, it is well known that there are no sources that can move the event horizon so that the marginally trapped surface $\mu$ lies outside if the matter corresponds to sources satisfying the NEC \cite{Hawking:1971tu}. However, $f(R)$ gravity differs from Einstein gravity in that the well-defined expansion satisfying the focusing theorem is the generalized expansion \eqref{definition of Generalized Expansion in section 1.2}, and the correspondence focusing theorem is
\begin{equation}
     k^{a}\nabla_{a}\Theta_{k}=-\frac{\theta_{k}^{2}}{D-2}-(k^{a}\nabla_{a}\text{log}f'(R))^{2}-\varsigma^{ab}\varsigma_{ab}-8\pi G\frac{T_{ab}k^{a}k^{b}}{f'(R)}.
     \label{section 6 focusing theorem}
\end{equation}
Therefore, it is natural to consider that: 
\begin{itemize}
    \item In $f(R)$ gravity, the event horizon cannot move so far inward such that generalized marginally trapped surface $\mu$ lies outside if the matter satisfies NEC.
\end{itemize}
Proof: Suppose there exists a generalized marginally trapped surface outside the event horizon. This implies that one can fire a light ray reaching the asymptotic AdS boundary; however, as the light ray approaches the boundary, its expansion will diverge. And this is in contradiction to the focusing theorem \eqref{section 6 focusing theorem}. Therefore, the generalized marginally trapped surface $\mu$ must lie behind the event horizon.

The above discussions show that the generalized marginally trapped surface $\mu$ cannot lie inside $R_{J}[t_{i}]$. This shows that
\begin{equation}
    R_{J}[t_{i}]\subseteq O^{f}_{W}[\mu].
\end{equation}
Thus, the one-point data $\langle O\rangle_{J}$ allows us to reconstruct at most $O^{f}_{W}[\mu]$, that is, the set of data used to compute the simple entropy is a subset of the set of data used to compute the outer entropy. This will lead to
\begin{equation}
    S^{\text{(simple)}}[t_{i}]\geq S^{\text{(outer)}}_{f}[\mu].
\end{equation}
Therefore, we show that the outer entropy is the lower bound of the simple entropy.

Finally, we are going to show that for the near equilibrium state, on the bulk side corresponding to near stationary black holes, the simple entropy coincides with the outer entropy of the generalized marginally trapped surface $\mu$. For thermal equilibrium state $\rho_{\text{stat}}$, the corresponding geometry is the stationary black hole in $f(R)$ gravity, denote as $\mathcal{M}_{\text{stat}}$. For a near thermal equilibrium state $\rho$ with simple source $J$ be turned on after time $t_{i}$, on the bulk side, we can ``turn off'' the matter falling into the event horizon by attaching a stationary null hypersurface $N_{k}[\mu]$, analogous to the procedure in section \ref{Construction of Stationary Null Hypersurface For the Generalized Expansion}. Using the HKLL prescription, we can then reconstruct the boundary simple source $J'$ for the new geometry $\mathcal{M}'$. In this geometry, the generalized marginally trapped surface $\mu$ is just on the event horizon\footnote{In the Einstein frame, for big perturbation from thermal equilibrium state, even we turn off all the ``matter'' (include the $f'(R)$ field, in the Einstein frame is the $\phi$) on event horizon, we are not guaranteed that $\mu$ lies on event horizon \cite{Engelhardt:2018kcs}. This is the reason why we only talk about near thermal equilibrium state.}, then we can fix all dynamical information on $N_{l}[t_{i}]$ outside $\mu$. Together with original source $J$, we can fully reconstruct all the data in the outer wedge $O^{f}_{W}[\mu]$, this shows that for the $f(R)$ gravity
\begin{equation}
    S^{\text{(simple)}}[t_{i}]=S^{\text{(outer)}}_{f}[\mu].
\end{equation}

\subsection{Discussion of the Second Law}

We first demonstrate that both the outer and simple entropy satisfy the second law. In the bulk, due to the $f(R)$ gravity focusing theorem \eqref{section 6 focusing theorem}, a later-time generalized marginally trapped surface $\mu_2$ (corresponding to boundary time $t_2$) must lie within the outer wedge of an earlier-time surface $\mu_1$ (corresponding to boundary time $t_{1}$), and naturally $O^{f}_{W}[\mu_{2}]\subset O^{f}_{W}[\mu_{1}]$. Therefore, there are fewer data in $O^{f}_{W}[\mu_{2}]$ compared to $O^{f}_{W}[\mu_{1}]$, and naturally
\begin{equation}
    S^{\text{(outer)}}_{f}[\mu_{2}]\geq S^{\text{(outer)}}_{f}[\mu_{1}].
    \label{second law of outer entropy}
\end{equation}

On the boundary side, things become more natural. Consider the simple entropy at two different times $t_{1}$ and $t_{2}$, where $t_{2}> t_{1}$. The boundary argument is analogous: since there are fewer constraints (regarding simple sources and one-point functions) after $t_2$ than after $t_1$, the entropy is monotonically non-decreasing. Therefore, the simple entropy satisfies the second law
\begin{equation}
    S^{\text{(simple)}}[t_{2}]\geq S^{\text{(simple)}}[t_{1}].
    \label{second law of simple entropy}
\end{equation}

For the near equilibrium black hole in $f(R)$ gravity, under the first order perturbation the dynamical entropy proposed by Hollands-Wald-Zhan \cite{Visser:2024pwz, Hollands:2024vbe}:
\begin{equation}
    S_{\text{dyn}}=(1-v\partial_{v})\frac{1}{4G}\int_{\mathcal{C}(v)} f'(R)=(1-v\partial_{v})S_{\text{Wald}}[\mathcal{C}(v)],
\end{equation}
here $v$ is the affine null parameter along the future horizon and $\mathcal{C}(v)$ is an arbitrary cross section. Under a first-order perturbation, it can be shown that the dynamical black hole entropy takes the form of the Wald entropy of the generalized marginally trapped surface \cite{Furugori:2025pmn, Kong:2024sqc, Jia:2025tgf}
\begin{equation}
    S_{\text{dyn}}(v)=\frac{1}{4G}\int_{\mu(v)}f'(R),
\end{equation}
here $\mu(v)$ is the associate generalized marginally trapped surface or the generalized apparent horizon of $\mathcal{C}(v)$. This agrees with our result \eqref{section 4 outer entropy final result} in the first order sense
\begin{equation}
    S_{\text{dyn}}(v)=S_{f}^{\text{(outer)}}[\mu(v)]=S^{\text{(simple)}}[t],
\end{equation}
here $t$ is the boundary time corresponding to $\mu(v)$. And if the matter satisfies the NEC, the dynamical black entropy satisfies the second law
\begin{equation}
    \partial_{v}\delta S_{\text{dyn}}=\frac{2\pi}{\kappa}\int_{\mathcal{C}(v)}\delta T_{ab}\xi^{a}k^{a}\geq 0,
\end{equation}
here $\kappa$ is the surface gravity and $\xi^{a}=\kappa vk^{a}$ is the killing field (after turning on falling matter $\xi^{a}$ will no longer be the killing field) on the future horizon. Hence, this result agrees with the \eqref{second law of outer entropy} and \eqref{second law of simple entropy}.

From the above discussions, we can therefore conclude that the outer entropy and its boundary dual, which is the simple entropy, are the correct holographic interpretation of the dynamical black hole entropy of $f(R)$ gravity, at least in the first order perturbation near stationary black hole in the large N limit where the correspondent bulk physics are classical gravity. In particular, our construction is even a non-perturbative construction, at least valid order by order in perturbation to a stationary black hole in $f(R)$ gravity, which can be viewed as a generalization of the dynamical black hole entropy of $f(R)$ gravity in AdS/CFT.

\section{Conclusion}

In this paper, we revisit and develop the holographic description of $S_{\rm dyn}$ in $f(R)$ gravity proposed by Hollands–Wald–Zhang \cite{Hollands:2024vbe}, placing it into a more systematic AdS/CFT framework and tracking carefully how the entropy dictionary is modified by higher-curvature dynamics. On the gravity (bulk) side, we establish that the outer entropy of a generalized marginally trapped surface is simply equal to the Wald entropy evaluated on that surface. In other words, once the “outer wedge” data are held fixed, the maximal coarse-grained entropy compatible with the exterior geometry is controlled by the same higher-curvature functional that governs black-hole entropy in $f(R)$ theories.

A key ingredient of our analysis is a careful treatment of Einstein frames and their holographic interpretation. We first derive holographic description in the Einstein frame, where the theory can be written as Einstein gravity coupled to an auxiliary scalar degree of freedom, and where standard holographic techniques apply  directly. Within this setup, we then establish an explicit correspondence between the boundary von Neumann entropy computed in the Einstein frame description and the corresponding entropy in the $f(R)$ frame. Concretely, we track how the reduced density matrix and its entropy transform under the field redefinitions and Weyl rescalings relating the two frames, and we identify the precise map that preserves the physical content of entanglement. Using this entropy dictionary, we translate the outer entropy construction between frames, obtaining a robust outer entropy correspondence that allows us to import the Einstein frame maximization logic and re-express it in purely geometric language in the $f(R)$ frame. This provides the conceptual bridge needed to equate the coarse-grained outer entropy with the Wald functional of the generalized marginally trapped surface.

Building on this foundation, we proceed to derive the outer entropy directly in $f(R)$ gravity. We prove a focusing theorem for an appropriately defined generalized expansion, which serves as a higher-curvature generalization of the Raychaudhuri equation under the usual null energy conditions. With this theorem in hand, we construct a stationary null hypersurface characterized by vanishing generalized expansion, and we formulate the necessary junction conditions for consistently stitching geometries across a hypersurface within $f(R)$ gravity. These tools allow us to explicitly construct a bulk spacetime whose boundary von Neumann entropy saturates the upper bound implied by the outer entropy maximization—i.e., we realize the extremal configuration achieving the coarse-grained maximum compatible with the fixed exterior. Finally, using the $f(R)$ focusing theorem, we show that the simple entropy defined on the boundary is indeed the holographic dual of the outer entropy in $f(R)$ gravity, thereby extending the outer entropy and simple entropy correspondence to this higher-curvature setting and show the second law of the two entropy which agree with the dynamical black hole entropy \cite{Hollands:2024vbe}.

Looking forward, our results suggest several promising future directions. Since our construction of the coarse graining entropy is non-perturbative in classical gravity (at least to all orders in perturbation theory around equilibrium), it suggests that the non-perturbative construction of the dynamical black hole entropy is highly related to the generalized marginally trapped surface (generalized minimar surface). In a recent paper \cite{Ashtekar:2025qqa} it was suggested that quasi-local horizon (QLH) is the right geometric concept for far from equilibrium black hole thermodynamics in Einstein gravity. We hope that similar concepts will appear in general higher curvature gravity. Recently there are some discussions about higher curvature focusing theorem near first order dynamic perturbative killing horizon \cite{Yan:2024gbz, Yan:2025vdp}, our construction of a non-perturbative focusing theorem in $f(R)$ gravity suggests similar theorem may exists in more general higher curvature gravity, for example, in $f(\text{Riemann})$ gravity. This will help us define HRT surface in general higher curvature gravity, and explain why the Wall entropy and Dong entropy \cite{Dong:2013qoa, Wall:2015raa} are the same but with different physical original.

\acknowledgments

We would like to thank Antony J. Speranza, Zihan Yan and Jiyu Cheng for useful discussions. In particular, we are greatly grateful to Antony J. Speranza for very patient discussions and useful comments on the draft. We are also grateful for the beautiful environment and academic atmosphere of University of Amsterdam, where this work was completed.

\appendix
\section{Cases of \texorpdfstring{$f''(R)=0$}{f''(R)=0}}
\label{Cases of f''(R)=0}

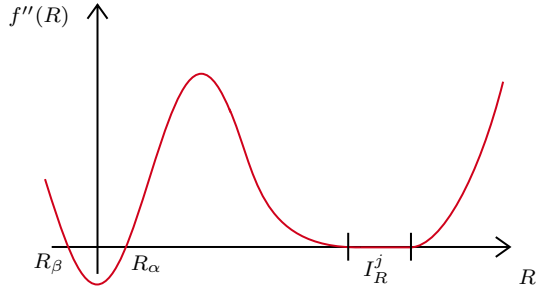
\begin{figure}[h]
    \centering

\tikzset{every picture/.style={line width=0.75pt}} 

\begin{tikzpicture}[x=0.75pt,y=0.75pt,yscale=-1,xscale=1]

\draw  (130.67,179.13) -- (359.33,179.13)(153.53,57.33) -- (153.53,192.67) (352.33,174.13) -- (359.33,179.13) -- (352.33,184.13) (148.53,64.33) -- (153.53,57.33) -- (158.53,64.33)  ;
\draw  [color={rgb, 255:red, 208; green, 2; blue, 27 }  ,draw opacity=1 ] (127.33,145) .. controls (135.81,172.15) and (143.92,198) .. (153.33,198) .. controls (162.74,198) and (170.85,172.15) .. (179.33,145) .. controls (187.81,117.85) and (195.92,92) .. (205.33,92) .. controls (210.52,92) and (215.32,99.87) .. (220,111.48) ;
\draw [color={rgb, 255:red, 208; green, 2; blue, 27 }  ,draw opacity=1 ]   (284.67,179.33) .. controls (232.67,180) and (232.67,140) .. (219.33,109.67) ;
\draw [color={rgb, 255:red, 208; green, 2; blue, 27 }  ,draw opacity=1 ]   (284.67,179.33) -- (310,179.33) ;
\draw [color={rgb, 255:red, 208; green, 2; blue, 27 }  ,draw opacity=1 ]   (310,179.33) .. controls (325.33,178.67) and (344,144) .. (356,96) ;
\draw [color={rgb, 255:red, 0; green, 0; blue, 0 }  ,draw opacity=1 ]   (278.67,172) -- (278.67,186) ;
\draw [color={rgb, 255:red, 0; green, 0; blue, 0 }  ,draw opacity=1 ]   (310,172) -- (310,186) ;

\draw (124.37,63.2) node  [font=\scriptsize]  {$f''( R)$};
\draw (368,193.87) node  [font=\scriptsize]  {$R$};
\draw (177.92,187.53) node  [font=\scriptsize]  {$R_{\alpha }$};
\draw (129.25,187.53) node  [font=\scriptsize]  {$R_{\beta }$};
\draw (292.58,190.2) node  [font=\scriptsize]  {$I_{R}^{j}$};

\end{tikzpicture}
\caption{This is an example of $f''(R)$. We use $R_{\alpha}$, $R_{\beta}$ denote the discrete zeros, $\alpha$ and $\beta$ are used to label different zeros. And use $I^{j}_{R}$ denote the interval zero, that is, if $R$ is belong to the interval $I^{j}_{R}$, then $f''(R)=0$. Here $j$ is used to label different interval zero.}
\label{f''(R) function}
\end{figure}

For points on the stationary null surface $N_{-k}$ satisfying $f''(R)\ne 0$, the condition $\partial_v R=0$ is ensured. We now consider regions on $N_{-k}$ where $f''(R)=0$. We first introduce relevant definitions. For the considered spacetime, the pre-spacetime regions for discrete zeros $R_\alpha$ (see figure \ref{f''(R) function}) are defined as the subregions of spacetime where the curvature equals $R_\alpha$, namely:
\begin{equation}
    \mathcal{M}_{R_{\alpha}}=\{p\in\mathcal{M}|R|_{p}=R_{\alpha}\}.
\end{equation}
Similarly, the pre-spacetime regions for interval zero $I^{j}_{R}$ (see figure \ref{f''(R) function}) are defined as the subregion of spacetime with curvature belonging to interval zero $I^{j}_{R}$
\begin{equation}
    \mathcal{M}_{I^{j}_{R}}=\{p\in\mathcal{M}|R|_{p}\in I^{j}_{R}\}.
\end{equation}

In the first case, we define $V_{R_\alpha}=N_{-k}\cap\mathcal{M}_{R_\alpha}$ and consider its $k^a$ congruence, denoted as $N_{\pm k}[V_{R_\alpha}]$. In the complement region of the congruence, that is, $N_{\pm k}[V_{R_{\alpha}}]/V_{R_{\alpha}}$, since $f''(R)\neq0$ we have $\partial_{v}R=0$. Assuming spacetime continuity, the curvature over the entire congruence must match that of $V_{R_\alpha}$. Therefore, this will imply that $\partial_{v}R|_{N_{\pm k}[V_{R_{\alpha}}]}=0$ and naturally $\partial_{v}R|_{V_{R_{\alpha}}}=0$.

For the other case, we define $V_{I^{j}_{R}}=N_{-k}\cap\mathcal{M}_{I^{j}_{R}}$. If the spacetime curvature falls within the interval $I_R^j$, the $f(R)$ gravity reduces to Einstein gravity, that is, $f(R)=cR+\text{constant}$. Together with the equation of motion of $f(R)$ gravity, we can show that
\begin{equation}
    \partial_{v}R\propto \partial_{v}T=\partial_{v}(g^{ab}T_{ab}).
\end{equation}
Since $g^{uu}=g^{ui}=g^{vi}=0$ on $N_{-k}$, we only need to consider $T_{uv}$ and $T_{ij}$. Imposing the initial free data conditions \eqref{initial free data Tij} and \eqref{initial free data Tuv} yields $\partial_v T=0$. Therefore, this will imply $\partial_{v}R|_{V_{I^{j}_{R}}}=0$. In conclusion, even if there exist subregions of $N_{-k}$ where $f''(R)=0$, the result $\partial_v R=0$ remains correct.

\bibliographystyle{JHEP}
\bibliography{biblio.bib}

\end{document}